\documentclass[preprint,3p,11pt,sort&compress]{elsarticle}
\usepackage{amssymb}
\usepackage{amsthm}
\usepackage{amsmath}
\usepackage{epstopdf}
\usepackage{subfigure}
\usepackage{longtable,booktabs}
\usepackage{color}
\usepackage{hyperref}
\usepackage{bm} % for \bm
\usepackage{pgfplots}
\pgfplotsset{compat=newest}
\usetikzlibrary{plotmarks}
\pgfplotsset{plot coordinates/math parser=false}
\newlength\figureheight 
\newlength\figurewidth
\tikzset{every picture/.append style={scale=0.8}}

\newcommand{\Eref}[1]{Equation (\ref{#1})}
\newcommand{\fref}[1]{Figure (\ref{#1})}

\usepackage{setspace}
\graphicspath{ {Figures/}}

\begin{document}

\begin{frontmatter}
 
\title{Isogeometric analysis of thin Reissner-Mindlin plates and shells: locking phenomena and  B-bar method}

\author[dut1,lu1]{Qingyuan Hu}
\ead{qingyuanhucn@gmail.com}

\author[dut2]{Yang Xia}

\author[iit]{Sundararajan Natarajan}

\author[lu1]{Andreas Zilian}

\author[dut2]{Ping Hu}

\author[dtu,lu1]{St\'ephane P.A. Bordas\corref{cor1}}
\ead{stephane.bordas@gmail.com}
\cortext[cor1]{Corresponding author}

\address[dut1]{Department of Engineering Mechanics, Dalian University of Technology, Dalian 116024, P.R. China}
%\address[lu1]{Institute of Computational Engineering, University of Luxembourg, L-4364 Esch-sur-Alzette, Luxembourg}
\address[lu1]{Department of Computational Engineering Sciences, Faculty of Sciences, Technology and Communication, University of Luxembourg, Luxembourg}
\address[dut2]{School of Automotive Engineering, Dalian University of Technology, Dalian 116024, P.R. China}
\address[iit]{Integrated Modelling and Simulation Lab, Department of Mechanical Engineering, Indian Institute of Technology, Madras, Chennai-600036, India}
\address[dtu]{Visiting Professor, Institute of Research and Development, Duy Tan University, K7/25 Quang Trung, Danang, Vietnam}

\begin{abstract}
We propose a local type of B-bar formulation,
addressing locking in degenerated Reissner-Mindlin plate and shell formulations in the context of isogeometric analysis.
Parasitic strain components are projected onto the physical space locally,
i.e. at the element level, using a least-squares approach.
The formulation allows the flexible utilization of basis functions of different order as the projection bases.
%
%To address the locking syndrome of degenerated Reissner-Mindlin plates and shells within the framework of isogeometric analysis, a generalized local $\bar{B}$ framework is proposed.
%Firstly, a local $\bar{B}$ method is implemented by projecting the locking strain locally, i.e. within each element using the $L_2$ norm.
%Moreover, the generalized form is proposed to utilize basis functions of different orders as the projected bases.
%
The present formulation is much cheaper computationally than the classical $\bar{B}$ method.
We show the numerical consistency of the scheme through numerical examples,
moreover they show that the proposed formulation alleviates locking and yields good accuracy 
even for slenderness ratios of $1 \times 10^5$,
and has the ability to capture deformations of thin shells using relatively coarse meshes.
In addition it can be opined that the proposed method is less sensitive to locking and mesh distortion.
%Our numerical results also indicate that the proposed method maintains good accuracy when both locking and mesh distortion are present at the same time.
\end{abstract}

\end{frontmatter}
%\linenumbers

% added by Qingyuan HU
\section*{Highlights}

$\bullet$ A local type of B-bar formulation is proposed to address the locking phenomenon in Reissner-Mindlin plates and shells.

$\bullet$ The parasitic strains are projected onto the physical space locally thus proposed formulation is computationally less expensive than classical B-bar method.

$\bullet$ Different sets of basis functions are used as the projection bases to achieve better performance.

$\bullet$ The formulation is less sensitive to mesh distortion when locking happens than pure IGA.

\section{Introduction}

The conventional Lagrange-based finite element method (FEM) employs polynomial basis functions to represent the geometry and the unknown fields.
The commonly employed approximation functions are Lagrangian polynomials.
However, these Lagrange polynomials are usually built upon a mesh structure which needs to be generated, from the CAD or Image file provided for the domain of interest. 
This mesh generation leads to the loss of certain geometrical features: e.g. a circle becomes a polyhedral domain. 
Moreover, Lagrange polynomials lead to low order continuity at the interface between elements, which is disadvantageous in applications requiring high order partial differential equations.

The introduction of isogeometric analysis (IGA)~\cite{hughes2005isogeometric} provides a general theoretical framework for the concept of ``design-through-analysis'' which has attracted considerable attention. The key idea of IGA is to provide a direct link between the computer aided design (CAD) and the simulation, by utilizing the same functions to approximate the unknown field variables as those used to describe the geometry of the domain under consideration, similar to the idea proposed in \cite{kagan1998new}. Moreover, it also provides a systematic construction of high-order basis functions~\cite{lipton2010robustness}. Note that, more recently, a generalisation of the isogeometric concept was proposed, whereby the geometry continues to be described by NURBS functions, as in the CAD, but the unknown field variables are allowed to live in different (spline) spaces. This lead to the concept of sub and super-geometric analysis, also known as Geometry Independent Field approximaTion (GIFT), described within a boundary element framework in \cite{marussig2015fast} and proposed in \cite{xu2014geometry,xu2014geometry2} and later refined in  \cite{atroshchenko2017weakening}. Related ideas, aiming at the construction of tailored spline spaces for local refinement were proposed recently in \cite{toshniwal2017smooth}.

In the literature, the IGA has been applied to study the response of plate and shell structures, involving two main theories, viz., the Kirchhoff-Love theory and the Reissner-Mindlin theory.
Thanks to the $\mathcal{C}^1$-continuity of the NURBS basis functions adopted in IGA,
Kiendl \textit{et al.}~\cite{kiendl2009isogeometric} developed an isogeometric shell element based on Kirchhoff-Love shell theory. The isogeometric Kirchhoff-Love  shell element for large deformations was presented in \cite{benson2011large}.
The isogeometric Reissner-Mindlin shell element was implemented in \cite{benson2010isogeometric}, including linear elastic and nonlinear elasto-plastic constitutive behavior.
The blended shell formulation was proposed to glue the Kirchhoff-Love structures with Reissner-Mindlin structures in \cite{benson2013blended}. In addition, the isogeometric Reissner-Mindlin shell formulation that is derived from the continuum theory was presented in~\cite{dornisch2013isogeometric}, in which the exact director vectors were used to improve accuracy. The solid shell was developed in \cite{hosseini2013isogeometric}, in this formulation the NURBS basis functions were used to construct the mid-surface and a linear Lagrange basis function was used to interpolate the thickness field.

The Kirchhoff-Love type elements are rotation-free and are only valid for thin structures. Due to the absence of rotational degrees of freedom (DoFs), special techniques are required to deal with the rotational boundary conditions~\cite{cottrell2006isogeometric,kiendl2009isogeometric,hu2017skew} and multi-patch connection \cite{kiendl2010bending}.
Theoretically, the Reissner-Mindlin theory is valid for both thick and thin structures,
however it is observed from the literature \cite{benson2010isogeometric,echter2010numerical} that both the FEM and the IGA approaches suffer from locking for thin structures when the kinematics is represented by Reissner-Mindlin theory, especially for lower order elements and coarse meshes. This has attracted engineers and mathematicians to develop robust elements that alleviates this pathology.
Adam et.al. proposed a family of concise and effective selective and reduced integration (SRI) \cite{hughes1978reduced} rules for beams \cite{adam2014improved}, plates and shells \cite{adam2015improved}, and non-linear shells using T-splines \cite{adam2015reduced} within the IGA framework.
Elguedj et. al. \cite{Elguedj_2008_BoverbarFover_2732_2762} presented $\bar{B}$ method and $\bar{F}$ method to handle nearly incompressible linear and non-linear problems.
The $\bar{B}$ method has been successfully applied to shear locking problems in curved beams~\cite{Bouclier_2012_Lockingfreeisogeometric_144_162}, two dimensional solid shells~\cite{bouclier2013development}, three-dimensional solid shells \cite{bouclier2013efficient} and in nonlinear solid shell formulation \cite{bouclier2015isogeometric}.
Echter and Bischoff~ \cite{echter2010numerical,Echter_2013_hierarchicfamilyisogeometric_170_180} employed the e discrete shear gap (DSG) method \cite{bletzinger2000unified} within the IGA framework to alleviate shear locking syndrome effectively.
Other approaches include twist Kirchhoff theory~\cite{brezzi2010new}, virtual element method~\cite{brezzi2013virtual}, collocation method \cite{da2012avoiding,auricchio2013locking}, simple first order shear deformation theory \cite{yin2014isogeometric}, and single variable method \cite{kiendl2015single,beirao2015locking}.
The above approaches have been employed with Lagrangian elements and IGA framework with varying order of success.

The works of Robin Bouclier, Thomas Elguedj and Alian Combescure \cite{bouclier2013development,bouclier2013efficient,bouclier2015isogeometric} focused on solid-shells in IGA and achieve good un-locking performance,
thus it is worthy to further test the performance of the $\bar{B}$ method for degenerated Reissner-Mindlin plates/shells within IGA framework.
This paper builds on \cite{hu2016order} for beam and rod structures using Timoshenko theory,
in order to alleviate the locking phenomena,
the locking strains are projected onto lower order physical space by the least square method.
The novel idea behind the formulation is to use multiple sets of basis functions to project the locking strains locally i.e. element-wise,
instead of projecting globally i.e. all over the patch.
The local projecting algorithm is inspired by the local $\bar{B}$ method \cite{bouclier2013efficient} and also by the work of local least square method \cite{mitchell2011method,govindjee2012convergence}.
These kind of formulations allow one to perform least-square projections locally,
thereby reducing the computational effort significantly.

The outline of this paper is as follows:
Section \ref{rmplate} gives an overview of Reissner-Mindlin theory for plates and shells.
In Section \ref{glm}, we present the novel approach, the local $\bar{B}$ method to alleviate the locking (both shear and membrane) problems encountered in thin structures whilst employing Reissner-Mindlin formulation.
The robustness, accuracy and the convergence properties are demonstrated with some benchmark examples in Section \ref{numex},
followed by concluding remarks in the last section.

%---- end introduction

\section{Isogeometric formulation of Reissner-Mindlin plates and shells} \label{rmplate}
\subsection{Reissner-Mindlin shell model}

\begin{figure}[htbp]
\centering
\def\svgwidth{0.7\columnwidth}
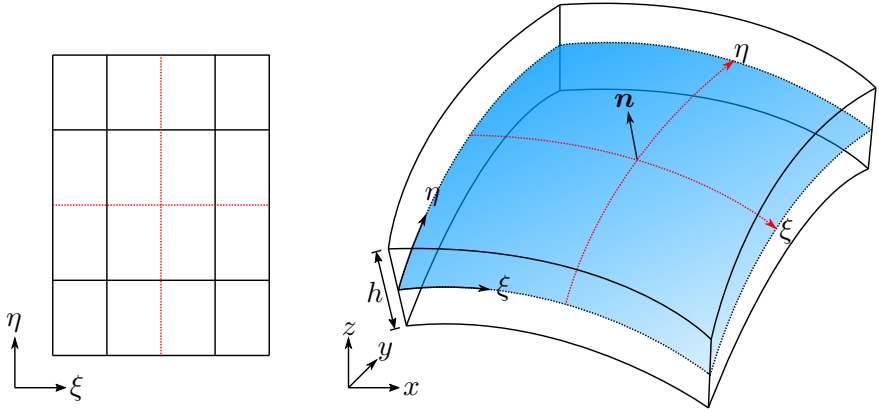
\caption{Mid-surface in the parameter space (left) and physical space (right) for a degenerated shell mid-surface. The real model is recovered by Eq.\eqref{geometry}.}
\label{shell}
\end{figure}

\fref{shell} represents the mid-surface of the shell in the parametric and physical spaces.
In IGA, the parametric space is typically Cartesian, while the physical domain of the undeformed shell can be of complex shape, not necessarily rectangular.
For simplicity, we consider a rectangular shell of constant thickness $h$,
and assume the linear elastic material to be homogeneous and isotropic,
which is described by Young's modulus $E$ and Poisson's ratio $\nu$.

The main difference between the Reissner-Mindlin and the Kirchhoff-Love shell theory is in the assumptions on the deformation behavior of the section and in the resulting independent kinematic quantities attached to the mid-surface in order to describe the deformation. 
According to the Reissner-Mindlin theory, a first order kinematic description is used in the thickness direction to account for the transverse shear deformations. Assuming a Cartesian coordinate system, any arbitrary point $P$ in the shell structure is described by:
\begin{equation}
\boldsymbol{x}_P=\boldsymbol{x}+\zeta \boldsymbol{n},
\label{geometry}
\end{equation}
and its displacement is calculated assuming small deformation as
\begin{equation}
\boldsymbol{u}_P=\boldsymbol{u}+\zeta \boldsymbol{\theta} \times \boldsymbol{n},
\label{displacement}
\end{equation}
where $\boldsymbol {x}$ is the geometry of the mid-surface as shown in Fig.\ref{shell},
$\zeta \in[-\frac{h}{2},\frac{h}{2}]$ denotes the thickness.
$\boldsymbol{u}$, $\boldsymbol{\theta}$ and $\boldsymbol{n}$ are the displacement vector, the rotation vector and the normal vector on the mid-surface point projected by point $P$.
The linearized strain tensor valid for small deformations is adopted here
\begin{equation}
\pmb{\varepsilon}=\frac{1}{2}(\boldsymbol{u}_{P,\boldsymbol{x}}+\boldsymbol{u}_{P,\boldsymbol{x}}^{\rm T}).
\label{strain}
\end{equation}

\subsection{Isogeometric approach}

In the context of shells, bi-variate NURBS basis functions are employed. Let $\Xi=\{\xi_1,\ldots,\xi_{n+p+1}\}$ and $H=\{\eta_1,\ldots,\eta_{m+q+1}\}$ be open knot vectors, and $w_A$ be given weights, $A=\{1,\ldots,nm\}$.
Then, the NURBS basis functions $R_A(\xi,\eta)$ are constructed, where $p$ and $q$ are the orders along the directions $\xi$ and $\eta$ respectively.
For more details about IGA, interested readers are referred to~\cite{nguyen2015isogeometric} and references therein.

Following the degenerated type formulation, the geometry of the undeformed mid-surface is described by
\begin{equation}
\boldsymbol{x}=\sum_{A=1}^{nm}R_A \boldsymbol{x}_A,
\label{geomt}
\end{equation}
and the mid-surface discrete displacement field is interpolated as
\begin{equation}
\boldsymbol{U^h}=\sum_{A=1}^{nm}R_A \boldsymbol{q}_A,
\label{cover}
\end{equation}
where $\boldsymbol{x}_A$ defines the location of the control points, $\boldsymbol{U^h}=(\boldsymbol{u},\boldsymbol{\theta})^{\rm T}$, $\boldsymbol{q}_A=(\boldsymbol{u}_A,\boldsymbol{\theta}_A)^{\rm T}$ is the vector of control variables corresponding to each control point, specifically $\boldsymbol{u}_A=(u,v,w)^{\rm T}$ and $\boldsymbol{\theta}_A=(\theta_x,\theta_y,\theta_z)^{\rm T}$.
The approximation space for displacement field is denoted as $Q_{p,q}$ in order to highlight the orders of the basis functions.

Once the mid-surface is described using Eq. \eqref{geomt} and Eq. \eqref{cover}, any arbitrary point $P$ in the shell body can be traced by the following discrete forms
\begin{equation}
\boldsymbol{x}_P=\sum_{A=1}^{nm}R_A(\boldsymbol{x}_A+\zeta \boldsymbol{n}_A),
\end{equation}
\begin{equation}
\boldsymbol{u}_P=\sum_{A=1}^{nm}R_A(\boldsymbol{u}_A+\zeta \boldsymbol{\theta}_A \times \boldsymbol{n}_A),
\end{equation}
where
\begin{equation}
\boldsymbol{n} = \frac{\boldsymbol{x}_{,\xi}\times\boldsymbol{x}_{,\eta}}{||\boldsymbol{x}_{,\xi}\times\boldsymbol{x}_{,\eta}||_2}
\end{equation}
is the normal vector.
The normal vectors at the Greville abscissae $\boldsymbol{n}_A$ are adopted here because it can achieve a good balance between the accuracy and the efficiency \cite{adam2015improved}.
It should be noted that the above equation includes the plate formulation, which can be considered as a special case.
For plates, one always has $\boldsymbol{n}(x,y,z)=(0,0,1)^{\rm T}$, which means that there are only two rotational degrees of freedom, $\theta_x$ and $\theta_y$.

Using Voigt notation, the relation between the strains and the stresses is expressed as
\begin{equation}
\boldsymbol{\sigma} =\boldsymbol{D}_g \pmb{\varepsilon},
\end{equation}
where $\boldsymbol{D}_g$ is the global constitutive matrix, and
\begin{equation}
\boldsymbol{D}_g =\boldsymbol{T}^{\rm T} \boldsymbol{D}_l \boldsymbol{T},
\end{equation}
here $\boldsymbol{D}_l$ is the given local constitutive matrix. To make $\boldsymbol{D}_l$ suitable for the physical geometry,
the transformation matrix $\boldsymbol{T}$ is employed, which is composed of $\boldsymbol{x}_{,\xi}$ and $\boldsymbol{x}_{,\eta}$.
In addition,
to fulfill the plane stress state $\sigma_{33}=0$,
the local constitutive matrix is given by
\begin{equation}
\boldsymbol{D}_l=
\frac{E}{1-\nu^2}
\begin{bmatrix}
1 & \nu & 0 & 0 & 0 & 0 \\
\nu & 1 & 0 & 0 & 0 & 0 \\
0 & 0 & 0 & 0 & 0 & 0 \\
0 & 0 & 0 & \frac{1-\nu}{2} & 0 & 0 \\
0 & 0 & 0 & 0 & \kappa\frac{1-\nu}{2} & 0 \\
0 & 0 & 0 & 0 & 0 & \kappa\frac{1-\nu}{2}
\end{bmatrix},
\end{equation}
in which $\kappa=\frac{5}{6}$ is the shear correction factor.

Upon employing the Galerkin framework and using the following discrete spaces for the displacement field,
\begin{align}
S&=\left\lbrace \boldsymbol{U}\in \left[ H^1(\varOmega)\right] ^d,\boldsymbol{U}|_{\varGamma_u}=\boldsymbol{U}^d\right\rbrace, \\
V&=\left\lbrace \boldsymbol{V}\in \left[ H^1(\varOmega)\right] ^d,\boldsymbol{V}|_{\varGamma_u}=0\right\rbrace, 
\end{align}
the variational function reads: find $\boldsymbol{U} \in S$ such that
\begin{equation}
b(\boldsymbol{U},\boldsymbol{U}^*)=l(\boldsymbol{U}^*)  \quad \forall \boldsymbol{U}^* \in V,
\end{equation}\
in which the bilinear term is
\begin{equation}
b(\boldsymbol{U},\boldsymbol{U}^*)=\int_{\varOmega}\pmb{\varepsilon}(\boldsymbol{U}^*)^{\rm T}\boldsymbol{D}_g \pmb{\varepsilon}(\boldsymbol{U})\rm d \varOmega.
\end{equation}

When the displacements and the rotations are approximated with polynomials from the same space, the discretized framework experiences locking (shear and membrane) when the thickness becomes very small.
The numerical procedure fails to satisfy the Kirchhoff limit as the shear strain does not vanish with the thickness of the shell approaching zero.
One explanation of shear locking is that different variables involved are not compatible \cite{kiendl2015single}, which is also known as field inconsistency.
Another explanation is that in curved elements shearless bending \cite{koschnick2005discrete} and inextensible bending deformations cannot be represented exactly, because of the appearance of spurious membrane and shear terms that absorb the major part of the strain energy \cite{Bouclier_2012_Lockingfreeisogeometric_144_162}, and this results in an overestimation of the stiffness.
In the next section, we introduce the B-bar method to deal with locking in Reissner-Mindlin plates and shells within the framework of IGA.

\section{B-bar method for Reissner-Mindlin plates and shells}
\label{glm}
In this section, 
we firstly illustrate the main idea behind the $\bar{B}$ method,
then after introducing the classical $\bar{B}$ method,
in order to further improve the efficiency of the formulation,
a local form of $\bar{B}$ formulation is proposed.

\subsection{The idea behind the B-bar method}

Imagining we have a unit length 2-node Timoshenko beam element at hand,
the element length parameter $x\in [0,1]$.
By the help of the shape functions
\begin{equation}
N_1 = 1-x, \quad N_2=x,
\end{equation}
the deflection field $w$ and the rotation field $\theta$ are built as
\begin{equation}
w = N_1w_1+N_2w_2, \quad \theta=N_1\theta_1+N_2\theta_2,
\end{equation}
in which the $w_i$ and $\theta_i$ are nodal variables, $i=1,2$.

In case of thin beams,
the Timoshenko beam theory tends to degenerate into the Euler beam theory,
and the shear strain (in Timoshenko beam theory) tends to vanish,
thus we have the following equation
\begin{equation}
\gamma = \frac{\rm d w}{\rm d x} - \theta = 0,
\end{equation}
and in discrete form
\begin{equation}\label{ieda_shear}
\gamma^h = (-w_1+w_2) - \big[ (1-x)\theta_1 + x\theta_2 \big] =0.
\end{equation}
The field inconsistency phenomenon occurs,
that is the order of $\theta_i$ is higher than the order of $w_i$,
which makes the shear strain difficult to vanish.

The idea of $\bar{B}$ method is to make the field to be order consistency.
We build a pseudo shear strain by a zero order basis functions as
\begin{equation}
\bar{\gamma}^h = \sum_{\bar{A}} \bar{N}_{\bar{A}} \bar{\gamma}^{\bar{A}^h}.
\end{equation}
Since the order of shape functions $N_i$ is one,
we define the one order lower set of shape functions as $\bar{N}_1=1$,
$\bar{A}=1$,
and $\bar{\gamma}^{1^h}$ is the corresponding pseudo DoF.
Next we perform a least square projection to solve for $\bar{\gamma}^{1^h}$
\begin{equation}
\int_0^1 \bar{N}_1 (\bar{\gamma}^h - \gamma^h) \, \rm d x = 0,
\end{equation}
we have
\begin{equation}
\bar{\gamma}^{1^h}+w_1-w_2+\frac{1}{2}\theta_1+\frac{1}{2}\theta_2 = 0.
\end{equation}
Solve the above equation for $\bar{\gamma}^{1^h}$
\begin{equation}
\bar{\gamma}^{1^h} = -w_1+w_2-\frac{1}{2}\theta_1-\frac{1}{2}\theta_2,
\end{equation}
and the pseudo shear strain we previously built becomes
\begin{equation}\label{idea_pseudo}
\bar{\gamma}^h = (-w_1+w_2)-( \frac{1}{2}\theta_1+\frac{1}{2}\theta_2 ).
\end{equation}
Compare the pseudo shear strain in \Eref{idea_pseudo} with the original one in \Eref{ieda_shear},
it is found that the shape functions for $w_i$ remain unchanged,
but the order of shape functions for $\theta_i$ is reduced to be the same as for $w_i$.
Thus it is believed that the pseudo shear strain could achieve a good un-locking performance.

\subsection{Classical B-bar method in IGA}
As stated before,
the novel idea behind the $\bar{B}$ method is using a modified strain instead of the original one.
In classical $\bar{B}$ method \cite{Elguedj_2008_BoverbarFover_2732_2762,Bouclier_2012_Lockingfreeisogeometric_144_162}, a common way is to use the  projection of the original strain to formulate the bilinear term
\begin{equation}
\bar{b}(\boldsymbol{U},\boldsymbol{U}^*)=\int_{\varOmega}\bar{\pmb{\varepsilon}}^{* \rm T} \boldsymbol{D} \bar{\pmb{\varepsilon}}\rm d \varOmega.
\end{equation}
The projected strain $\bar{\pmb{\varepsilon}}$ and the original strain $\pmb{\varepsilon}$ are equal in the sense of the least square projection.
The projection space is chosen to be one order lower, i.e. $Q_{\bar{p},\bar{q}}=Q_{p-1,q-1}$ (see \fref{strategy}(a)). Built from one order lower knot vectors, with all the weights given as $W_{\bar{A}}=1,\bar{A}=\left\lbrace 1,\ldots,\bar{n}\bar{m}\right\rbrace $, one order lower B-spline basis functions $\bar{N}_{\bar{A}}$ are obtained.
The $L_2$ projection process is performed on the physical domain as
\begin{equation}\label{ls}
\begin{split}
& \int_{\varOmega}\bar{N}_{\bar{B}}\left( \pmb{\varepsilon}^h-\bar{\pmb{\varepsilon}}^h \right) {\rm d}\varOmega=0, \\ 
& \bar{B}=1,\ldots,\bar{n}\bar{m},
\end{split}
\end{equation}
in which the discretized form of the projected strain is
\begin{equation}
\bar{\pmb{\varepsilon}}^h=\sum_{\bar{A}=1}^{\bar{n}\bar{m}}\bar{N}_{\bar{A}}\bar{\pmb{\varepsilon}}^{\bar{A}^h},
\end{equation}
where $\bar{\pmb{\varepsilon}}^{\bar{A}^h}$ means the projection of $\pmb{\varepsilon}^h$ onto $\bar{N}_{\bar{A}}$.
Finally, we have
\begin{equation}\label{final}
\bar{\pmb{\varepsilon}}^h=\sum_{\bar{A},\bar{B}=1}^{\bar{n}\bar{m}}\bar{N}_{\bar{A}}\boldsymbol{M}_{\bar{A}\bar{B}}^{-1}\int_{\varOmega}\pmb{\varepsilon}^h\bar{N}_{\bar{B}}{\rm d}\varOmega,
\end{equation}
where $\boldsymbol{M}_{\bar{A}\bar{B}}$ is is the inner product matrix
\begin{equation}
\boldsymbol{M}_{\bar{A}\bar{B}}=(\bar{N}_{\bar{A}},\bar{N}_{\bar{B}})_{\varOmega}=\int_{\varOmega}\bar{N}_{\bar{A}}\bar{N}_{\bar{B}}{\rm d}\varOmega.
\end{equation}

The Hu-Washizu principle could be utilized to prove the variational consistency of the $\bar{B}$ method \cite{simo1986variational,Elguedj_2008_BoverbarFover_2732_2762,Bouclier_2012_Lockingfreeisogeometric_144_162,bouclier2013development}.
In addition, \cite{Bouclier_2012_Lockingfreeisogeometric_144_162,bouclier2013efficient} proved that the $\bar{B}$ method is equivalent to the mixed formulation.

\subsection{Local B-bar formulation for Reissner-Mindlin plates and shells}

For degenerated plates and shells, although the geometries are represented in two dimensions,
and there is the thickness parameter $\zeta$ involved in the formulation.
In this case, if the whole part of strain is projected, rank deficiency appears and the formulation yields inaccurate results as shown in \cite{bouclier2013development} and also to our computational experience.
Following the similar approach as outlined in \cite{bouclier2013development,bouclier2013efficient}, in this work, only the average strain through the thickness is projected as
\begin{equation}
\overline{\boldsymbol{M\!I\!D}(\pmb{\varepsilon}_e^{h})}=\sum_{\bar{A},\bar{B}=1}^{(\bar{p}+1)(\bar{q}+1)}\bar{N}_{\bar{A}}\boldsymbol{M}_{\bar{A}\bar{B}}^{-1}\int_{\varOmega_e}\boldsymbol{M\!I\!D}(\pmb{\varepsilon}_e^{h})\bar{N}_{\bar{B}}{\rm d}\varOmega,
\end{equation}
where $\boldsymbol{M\!I\!D}(\pmb{\varepsilon}_e^{h})$ is defined to be the average strain through the thickness within a single element
\begin{equation}\label{MID}
\boldsymbol{M\!I\!D}(\pmb{\varepsilon}_e^{h})=\frac{1}{h}\int_{-\frac{h}{2}}^{\frac{h}{2}}\pmb{\varepsilon}_e^h{\rm d}z.
\end{equation}
The modified bi-linear form is defined as 
\begin{equation}
\bar{b}(\boldsymbol{U},\boldsymbol{U}^*)
=\int_{\varOmega}
\left(
\underbrace{ \pmb{\varepsilon}^{*{\rm T}}\boldsymbol{D}\pmb{\varepsilon} }_{\text{original}}
- 
\underbrace{ \boldsymbol{M\!I\!D}(\pmb{\varepsilon}^*)^{\rm T} \boldsymbol{D} \boldsymbol{M\!I\!D}(\pmb{\varepsilon}) }_{\text{original average strain}}
+ 
\underbrace{ \overline{\boldsymbol{M\!I\!D}(\pmb{\varepsilon})}^{\rm T} \boldsymbol{D} \overline{\boldsymbol{M\!I\!D}(\pmb{\varepsilon})} }_{\text{projected average strain}}
\right) {\rm d} \varOmega.
\end{equation}

Within local $\bar{B}$, if the shape functions over the elements possess $\mathcal{C}^0$ continuity (which is obviously fulfilled), the Hu-Washizu principle can be rewritten as \cite{huang1994quasi}
\begin{equation}
\delta\varPi_{HW}(\boldsymbol{U},\tilde{\pmb{\varepsilon}},\tilde{\boldsymbol{\sigma}})=\sum_m \left( \int_{\varOmega_m}\delta\tilde{\pmb{\varepsilon}}^{\rm T}\boldsymbol{D}\pmb{\varepsilon}{\rm d}\varOmega+\delta\int_{\varOmega_m}\tilde{\boldsymbol{\sigma}}^{\rm T}(\pmb{\varepsilon}-\tilde{\pmb{\varepsilon}}){\rm d}\varOmega-\int_{\varGamma_{fm}}\boldsymbol{t}\delta\boldsymbol{U}{\rm d}\varGamma_f \right) 
\end{equation}
in which $m$ denotes the number of elements. In this context the assumed displacements should satisfy $\mathcal{C}^0$ continuous between elements, but discontinuous assumed strains are allowed, this explains why different sets of basis functions can be used.
However, when the local $\bar{B}$ method is applied for degenerated plates,
i.e. the average strain is projected,
only the numerical consistency condition is satisfied according to the appendix of \cite{bouclier2013development}.
This conclusion also stands for the proposed method in this paper. 

Reviewing the literature of the $\bar{B}$ method since its appearance \cite{Hughes1980}, one valuable contribution is the introduction of local $\bar{B}$ concept \cite{bouclier2013efficient,hu2016order}, thanks to which lots of computational effort has been saved without too much accuracy loss.
The motivation of the local $\bar{B}$ concept comes from the observation that one needs to calculate the inverse of matrix $\boldsymbol{M}$ in \Eref{final}, which could be computationally expensive if the projection is applied globally. Thus, from a practical point of view, it is highly recommended to project the strains \emph{locally} \cite{bouclier2013efficient,hu2016order}, i.e. element-wise
\begin{equation}\label{ebe}
\bar{\pmb{\varepsilon}}_e^h=\sum_{\bar{A},\bar{B}=1}^{(\bar{p}+1)(\bar{q}+1)}\bar{N}_{\bar{A}}\boldsymbol{M}_{\bar{A}\bar{B}}^{-1}\int_{\varOmega_e}\pmb{\varepsilon}_e^h\bar{N}_{\bar{B}}{\rm d}\varOmega,
\end{equation}
with
\begin{equation}
\boldsymbol{M}_{\bar{A}\bar{B}}=(\bar{N}_{\bar{A}},\bar{N}_{\bar{B}})_{\varOmega_e}=\int_{\varOmega_e}\bar{N}_{\bar{A}}\bar{N}_{\bar{B}}{\rm d}\varOmega.
\end{equation}

\begin{figure}[htbp]
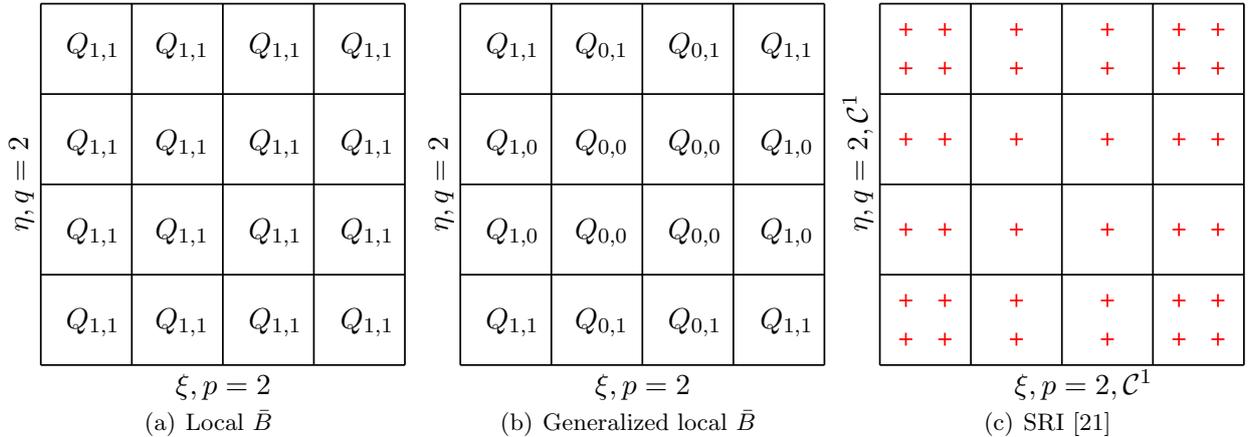

	\centering
    \def\svgwidth{0.32 \columnwidth}
	\subfigure[Local $\bar{B}$]{\input{strategy_LB.pdf_tex}}
    \def\svgwidth{0.32 \columnwidth}
	\subfigure[Generalized local $\bar{B}$]{\input{strategy_GLB.pdf_tex}}
    \def\svgwidth{0.32 \columnwidth}
	\subfigure[SRI \cite{adam2015improved}]{\input{strategy_SRI.pdf_tex}}
	\caption{Unlock strategies in bi-dimensional parameter space. Projecting spaces for locking strains are denoted as $Q_{p,q}$. For local $\bar{B}$ method (a), one order lower B-spline basis functions are employed for all elements. The presented generalized local $\bar{B}$ method (b) uses basis functions of different orders, inspired by the SRI method (c), while $\mathcal{C}^1$ continuities of elements are not restricted as in (c).}
	\label{strategy}
\end{figure}

Moreover, with the opinion of projecting the original strains into the lower order space could release the locking constrains,
we treat the corner, boundary and inner elements separately by using the lowest possible order of each element,
in order to release the locking constrains as much as possible.
Thus instead of projecting the strains onto $Q_{p-1,q-1}$, different sets of projection spaces are adopted in this work,
which could bring more flexibility into the formulation.
This is called the \emph{generalized strategy}.
The projection spaces need to be chosen carefully to avoid ill-condition or rand deficiency, readers interested in the corresponding mathematical theory is recommended to see \cite{antolin2016isogeometric}, which is in the context of volume locking (nearly-incompressible) problems.
Here, the strategy in our previous work \cite{hu2016order} is extended to bi-dimensional cases.
Specifically, for space $Q_{2,2}$, we adopt $Q_{1,1}$ for corner elements, $Q_{0,1}$ and $Q_{1,0}$ for boundary elements, and $Q_{0,0}$ for inner elements, as shown in \fref{strategy}(b).

\subsection{Discussions}

\begin{figure}[htbp]
	\centering
	\subfigure[Local $\bar{B}$]{\includegraphics[width=0.45\textwidth]{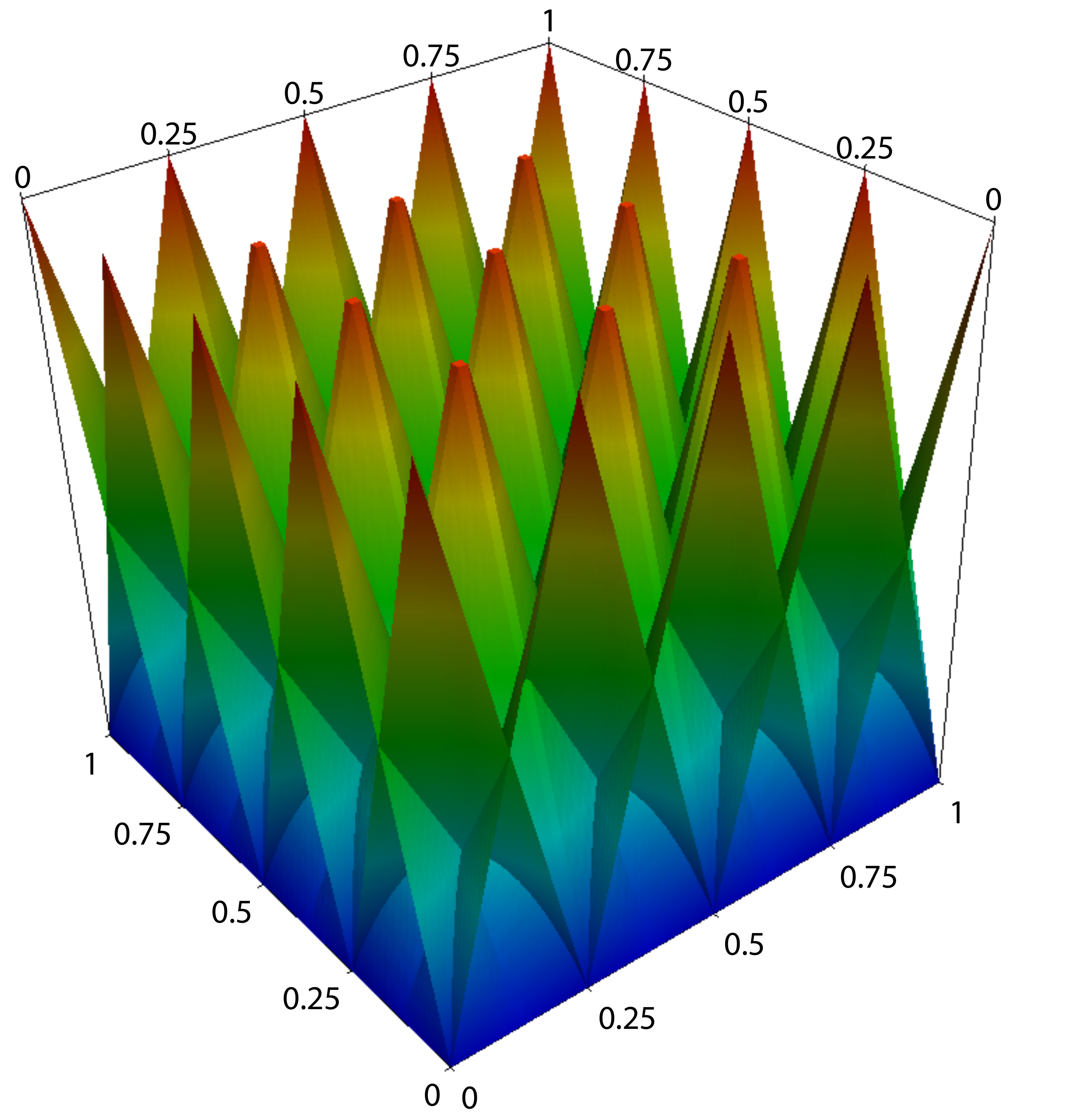}}
	\subfigure[Generalized local $\bar{B}$]{\includegraphics[width=0.45\textwidth]{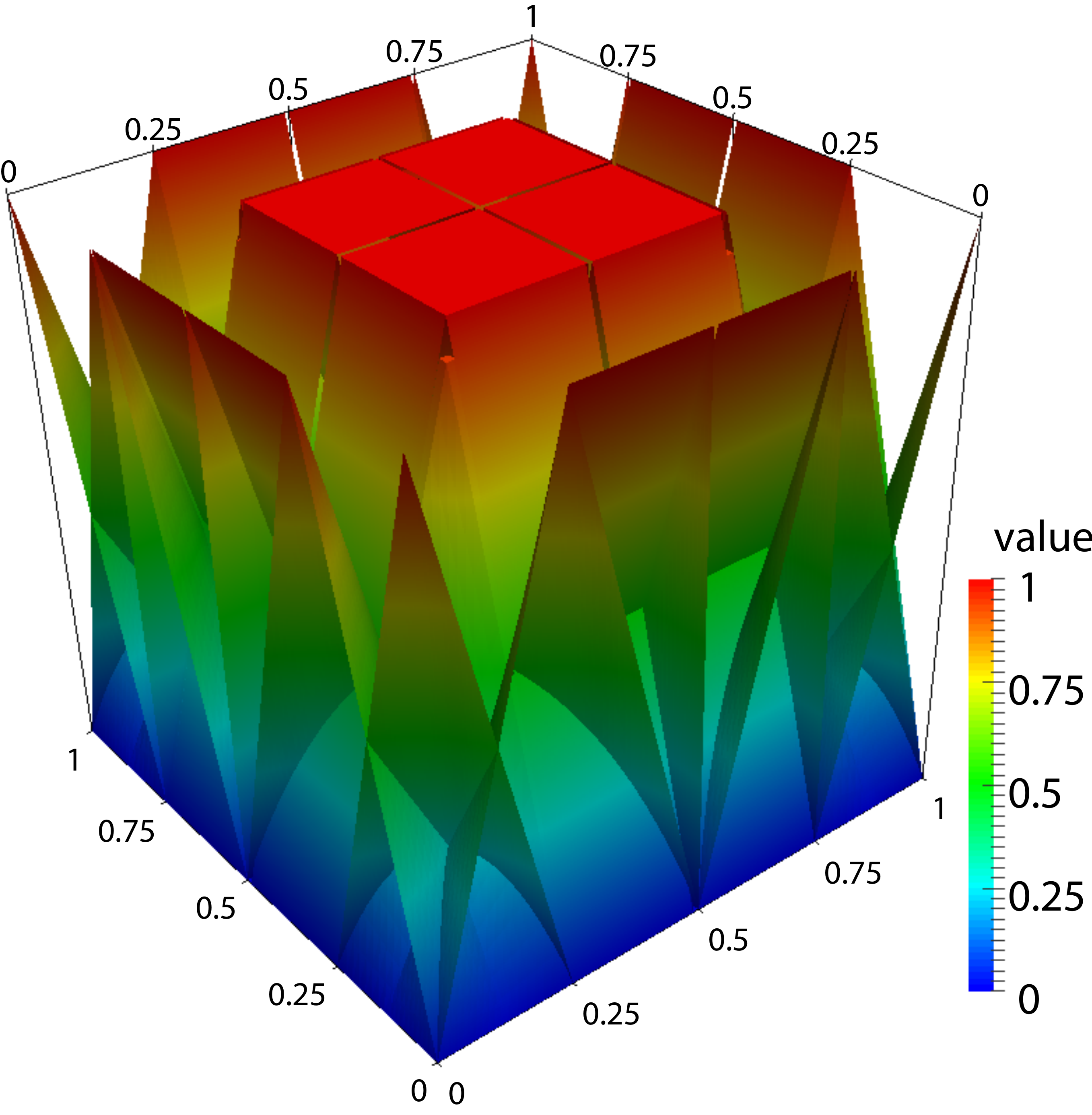}}
	\caption{Bi-dimensional basis functions for projecting locking strains of each element, let $Q_{p,q}$ denote the projecting space. There are $4\times 4$ elements. (a) Basis functions of $Q_{1,1}$ in the local $\bar{B}$ method. (b) Basis functions in $Q_{\bar{p},\bar{q}}$ (see Fig.\ref{strategy}) in the generalized local $\bar{B}$ method, the strains at the corners are assumed to be linear along both sides, the strains at the patch sides are assumed to be linear toward inside and constant along the boundary, the strains within the patch are assumed to be constant.}
	\label{plate_basis}
\end{figure}

The approach of using multiple sets of lower order basis functions was firstly presented for one-dimensional cases in \cite{hu2016order}.
This generalized strategy is similar with the SRI strategy in \cite{adam2015improved}.
In the case of only one quadrature point being used for an element, the functions are detected only at this quadrature point but nowhere else, which is analogous to its projection onto a $Q_{0,0}$ space.
To achieve a better understanding of the proposed projection strategy, the basis functions are shown in Fig.\ref{plate_basis}.
There are four elements per side.
In the local $\bar{B}$ method, only one single set of basis functions (i.e. $Q_{1,1}$) is used to form the projection space.
While in the generalized local $\bar{B}$ method, four set of basis functions (i.e. $Q_{\bar{p},\bar{q}}$) are used.
As shown in Fig.\ref{plate_basis} (b), this strategy means that the modified strains are assumed to be constant for the four internal elements and bi-linear for the four corner elements.
In particular, for the side elements, the modified strains are assumed to be constant along the patch boundary, and linear facing inside.

The advantage of SRI method is that its implementation is simple and the computations involve less calculations because fewer quadrature points are employed.
But one must consider the continuity between neighboring elements, which requires additional efforts.
However, for the local $\bar{B}$ method, there is no problem in continuity because the projection procedure is applied element-wise.
This conclusion can be found in \cite{hu2016order}.
Thus, one could always use knot vectors of order $\bar{p}$ and continuity $C^{\bar{p}-1}$ (i.e. without inner repeated knots) as the projection knot vector.
Apart from being accurate, since the lower order basis functions are used, for example linear functions for corner elements and constants for inner elements, better efficiency can be achieved with fewer quadrature points than usual.

Compared with the classical $\bar{B}$ method, the local $\bar{B}$ method shows promising advantages in terms of computational efficiency.
In the classical $\bar{B}$ method, the calculation of the inverse of matrix $\boldsymbol{M}$ requires large amounts of memory.
Thus, the adoption of local projection rather than global one saves lots of computational efforts.
Similar conclusions had been carried out in the context of LLSQ fitting for boundary conditions \cite{mitchell2011method} and further LLSQ algorithm \cite{govindjee2012convergence}. In these contributions, assuming $\mathbf{A}$ is the global Boolean assembly operator, the matrix
\begin{equation}
\left( \mathbf{A}\mathbf{A}^{\rm T}\right) ^{-1}\mathbf{A}
\label{strain_smooth}
\end{equation}
is used to calculate the uniformly weighted average of shared nodes. For instance, if a node (control point) is shared by $n$ elements, then the values corresponding to this node is divided by $n$.
Furthermore, for the local $\bar{B}$ method in \cite{bouclier2013efficient}, the same procedure named by strain smoothing is employed to ensure the continuity of the projected strains and thus to obtain results of better accuracy.
The price to pay is that one need to calculate the average operator and the bandwidth is larger than classical IGA.
In this research to obtain the strain field of higher order continuity, we use the original strain of discretized form $\varepsilon^h$ instead of $\bar{\varepsilon}^h$, i.e. recover the displacement field firstly by Eq.\eqref{cover} and then get the strain field as usual, in this way the continuity property of NURBS basis functions is utilized.

\section{Numerical examples}\label{numex}

In this section, we demonstrate the performance of the proposed generalized local $\bar{B}$ formulation for Reissner-Mindlin plates/shells by solving a few standard benchmark problems.
The proposed formulation is implemented within the open source C\texttt{++} IGA framework \emph{Gismo} \footnote{https://ricamsvn.ricam.oeaw.ac.at/trac/gismo/wiki/WikiStart} \cite{jlmmz2014}.
The numerical examples include: (a) Square plate; (b) Scordelis-Lo roof; (c) Pinched cylinder and (d) Pinched hemisphere with a hole.
Unless otherwise mentioned consistent units are employed in this study.
In all the numerical examples, the following knot vectors are chosen as the initial ones:
\begin{equation}
\begin{split}
\Xi=\{0,0,0,1,1,1\}, \\
H=\{0,0,0,1,1,1\}.
\end{split}
\end{equation}
In the following examples,
$(p+1)\times(q+1)$ Gauss points are used for numerical integration,
$p\times q$ Gauss points are used for the family of $\bar{B}$ methods,
the reduced quadrature scheme in \cite{adam2015improved} is adopted for comparison.
The following conventions are employed whilst discussing the results:
\begin{itemize}
\item LB:\ Local $\bar{B}$ method \cite{bouclier2013efficient} without strain smoothing in \Eref{strain_smooth}. The projection is applied element by element as shown in Eq.\eqref{ebe} and \fref{strategy}(a),
\item GLB:\ Generalized local $\bar{B}$ method, which means that based on LB, the strategy of using multi-sets of basis functions as shown in \fref{strategy}(b) is adopted,
\item SRI: \ Selective reduced integration \cite{adam2015improved}.
\end{itemize}

\subsection{Simply supported square plate \label{section41}}

\begin{figure}[htbp]
\centering
\def\svgwidth{0.7\columnwidth}
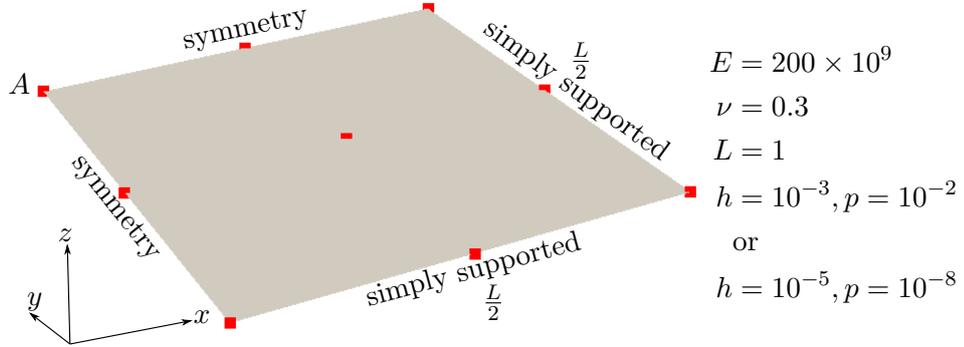
\caption{Mid-surface of the rectangular plate in Section \ref{section41}: geometry and boundary conditions. Only one quarter of the plate is shown here. The red filled squares are the corresponding control points.}
\label{plate}
\end{figure}

Consider a square plate simply supported on its edges with thickness $h$ and the length of the side $L=$ 1.
Owing to symmetry, only one quarter of the plate, i.e, $a=L/2$ is modeled as shown in \fref{plate}. 
The plate is assumed to be made up of homogeneous isotropic material with Young's modulus $E=$ 200 GPa and Poisson's ratio $\nu=$ 0.3,
and subjected to uniform pressure $p$.
The control points and the corresponding weights are given in Table \ref{geo_plate}.

\begin{table}[htbp]
%\scriptsize
\centering
\caption{Control points and weights for the rectangular plate in \fref{plate}.}
\begin{tabular}{crrrrrrrrr}
\hline\noalign{\smallskip}
 & 1 & 2 & 3&4&5&6& 7&8&9 \\	
 \hline		
$x$ & 0.5 & 0.75 & 1 & 0.5 & 0.75 & 1 & 0.5 & 0.75 & 1 \\
$y$ & 0 & 0 & 0 & 0.25 & 0.25 & 0.25 & 0.5 & 0.5 & 0.5 \\
$z$ & 0 & 0 & 0 & 0 & 0 & 0 & 0 & 0 & 0 \\
$w$ & 1 & 1 & 1&1&1&1&1&1&1 \\
\hline
\end{tabular}
\label{geo_plate}
\end{table}

The analytical out-of-plane displacements for a thin plate with simply supported edges are given by
\begin{equation}
w(x,y) =\frac{16p}{\pi^6 D} \sum\limits_{m=1,3,5,\cdots}^\infty \sum\limits_{n=1,3,5,\cdots}^\infty \frac{ \sin \left( \frac{m\pi x}{L} \right) \sin \left( \frac{n \pi y}{L} \right)} { mn\left[ \left( \frac{m}{L} \right)^2 + \left( \frac{n}{L} \right)^2 \right]^2}
\label{wK}
\end{equation}
where $D = \frac{E h^3}{12(1-\nu^2)}$.
The transverse displacement is constant when the applied load is proportional to $h^3$, thus the numerical solution does not depend on the thickness of the plate.
Due to the fact that the error of $w_A$ can express the field error to some extent,
only the errors of $w_A$ are studied instead of the field error,
the reference value is $w_A=$ -2.21804 $\times$ 10$^{-6}$.

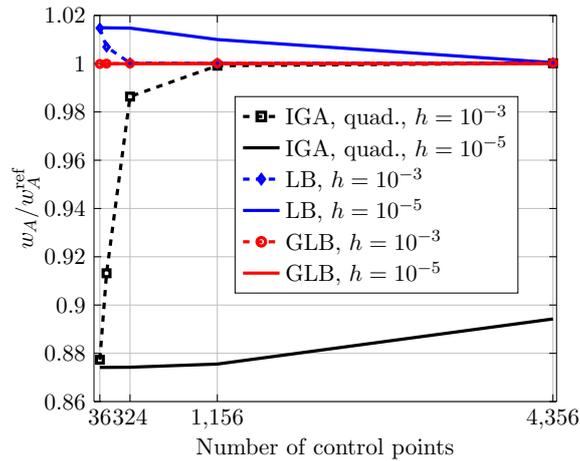
\begin{figure}[htpb]
\centering
\begin{tikzpicture}

\begin{axis}[%
width=0.951\figurewidth,
height=\figureheight,
at={(0\figurewidth,0\figureheight)},
scale only axis,
separate axis lines,
every outer x axis line/.append style={black},
every x tick label/.append style={font=\color{black}},
every x tick/.append style={black},
xmin=-20,
xmax=4390,
xtick={36,324,1156,4356},
xminorticks=true,
xlabel={Number of control points},
every outer y axis line/.append style={black},
every y tick label/.append style={font=\color{black}},
every y tick/.append style={black},
ymin=0.86,
ymax=1.02,
yminorticks=true,
ylabel={$w_A/w_A^{\text{ref}}$},
axis background/.style={fill=white},
xmajorgrids,
xminorgrids,
ymajorgrids,
yminorgrids,
legend style={at={(0.38,0.35)}, anchor=south west, legend cell align=left, align=left, draw=black}
]
\addplot [color=black, dashed, line width=1.5pt, mark=square, mark options={solid, black}]
  table[row sep=crcr]{%
%16	0.874364784\\
36	0.87733587\\
100	0.913164727\\
324	0.986368849\\
1156	0.999154886\\
4356 1.000169293\\
};
\addlegendentry{IGA, quad., $h=10^{-3}$}

\addplot [color=black, line width=1.5pt]
  table[row sep=crcr]{%
%16	0.874157394\\
36	0.874157394\\
100	0.874161903\\
324	0.87422953\\
1156	0.875546005\\
4356 0.894233637\\
};
\addlegendentry{IGA, quad., $h=10^{-5}$}

\addplot [color=blue, dashed, line width=1.5pt, mark=diamond, mark options={solid, blue}]
  table[row sep=crcr]{%
%16	1.01626906\\
36	1.014727162\\
100	1.006932007\\
324	1.000263971\\
1156	1.000142242\\
4356 1.000263971\\
};
\addlegendentry{LB, $h=10^{-3}$}

\addplot [color=blue, line width=1.5pt]
  table[row sep=crcr]{%
%16	1.016255535\\
36	1.014943569\\
100	1.014857907\\
324	1.014736179\\
1156	1.009993262\\
4356 1.000457836\\
};
\addlegendentry{LB, $h=10^{-5}$}

\addplot [color=red, dashed, line width=1.5pt, mark=o, mark options={solid, red}]
  table[row sep=crcr]{%
%16	1.01626906\\
36	0.999885259\\
100	1.000016005\\
324	1.000074615\\
1156	1.000142242\\
4356 1.000272988\\
};
\addlegendentry{GLB, $h=10^{-3}$}

\addplot [color=red, line width=1.5pt]
  table[row sep=crcr]{%
%16	1.016255535\\
36	0.999862717\\
100	0.999979937\\
324	1.00000248\\
1156	1.000011497\\
4356 1.000016005\\
};
\addlegendentry{GLB, $h=10^{-5}$}

\end{axis}
\end{tikzpicture}%
\caption{Normalized central displacement $w_A$ with mesh refinement for the rectangular plate \ref{plate}. $h$ stands for thickness. IGA of order 2 suffer from locking. GLB gets good accuracy.}
\label{lock_plate}
\end{figure}

\fref{lock_plate} shows the normalized displacement as a function of mesh refinement for two different plate thickness.
For thickness $h=$ 10$^{-3}$, although the conventional IGA yields inaccurate results for coarse meshes, the results tend to improve upon refinement.
However, it suffers from severe shear locking syndrome when $h=$ 10$^{-5}$.
In this case the results seem remain horizontal with respect to number of control points, indicating that the elements are fully locked.
For LB and GLB slight rank deficiency occur for coarse meshes.
GLB get quite good results when $h=$ 10 $^{-5}$.
From the numerical study it is inferred that the proposed formulation  alleviates shear locking phenomenon and yields accurate results even for extremely thin plates of slenderness ratio $10^5$. 

\begin{figure}[htpb]
\centering
\subfigure[Convergence]{% This file was created by matlab2tikz.
\definecolor{mycolor1}{rgb}{1.00000,0.00000,1.00000}%
\begin{tikzpicture}

\begin{axis}[%
width=0.951\figurewidth,
height=\figureheight,
at={(0\figurewidth,0\figureheight)},
scale only axis,
separate axis lines,
every outer x axis line/.append style={black},
every x tick label/.append style={font=\color{black}},
every x tick/.append style={black},
xmode=log,
xmin=1,
xmax=100,
xminorticks=true,
xlabel={Number of elements per side},
every outer y axis line/.append style={black},
every y tick label/.append style={font=\color{black}},
every y tick/.append style={black},
ymode=log,
ymin=1e-06,
ymax=1,
yminorticks=true,
ylabel={$|w_A-w_A^{\text{ref}}| / |w_A^{\text{ref}}|$},
axis background/.style={fill=white},
xmajorgrids,
xminorgrids,
ymajorgrids,
yminorgrids,
legend style={at={(0.02,-0.24)}, anchor=south west, legend cell align=left, align=left, fill=white}
]
\addplot [color=black, line width=1.5pt, mark=square, mark options={solid, black}]
  table[row sep=crcr]{%
2	1.25843E-01\\
4	1.25843E-01\\
8	1.25838E-01\\
16	1.25770E-01\\
32	1.24454E-01\\
64	1.05766E-01\\
};
\addlegendentry{IGA, quad.}

\addplot [color=blue, line width=1.5pt, mark=diamond, mark options={solid, blue}]
  table[row sep=crcr]{%
2	1.62555E-02\\
4	1.49436E-02\\
8	1.48579E-02\\
16	1.47362E-02\\
32	9.99326E-03\\
64	4.57836E-04\\
};
\addlegendentry{LB}

\addplot [color=red, line width=1.5pt, mark=o, mark options={solid, red}]
  table[row sep=crcr]{%
2	1.62555E-02\\
4	1.37283E-04\\
8	2.00627E-05\\
16	2.47966E-06\\
32	1.14966E-05\\
64	1.60051E-05\\
};
\addlegendentry{GLB}

\addplot [color=mycolor1, line width=1.5pt, mark=triangle, mark options={solid, mycolor1}]
  table[row sep=crcr]{%
2	1.48579E-02\\
4	1.48579E-02\\
8	1.48263E-02\\
16	1.31808E-02\\
32	1.46323E-03\\
64	1.82433E-02\\
};
\addlegendentry{IGA, cubic}

\end{axis}

%\node at (4.4,0.38) {1.39};

\end{tikzpicture}%
}
\subfigure[Partical enlarge]{% This file was created by matlab2tikz.
\definecolor{mycolor1}{rgb}{1.00000,0.00000,1.00000}%
\begin{tikzpicture}

\begin{axis}[%
width=\figurewidth,
height=\figureheight,
at={(0\figurewidth,0\figureheight)},
scale only axis,
unbounded coords=jump,
separate axis lines,
every outer x axis line/.append style={black},
every x tick label/.append style={font=\color{black}},
every x tick/.append style={black},
xmode=log,
xmin=1,
xmax=100,
xminorticks=true,
xlabel={Number of elements per side},
every outer y axis line/.append style={black},
every y tick label/.append style={font=\color{black}},
every y tick/.append style={black},
ymin=-2.28e-6,
ymax=-2.16e-6,
y dir=reverse,
ylabel={$w_A$},
axis background/.style={fill=white},
xmajorgrids,
xminorgrids,
ymajorgrids,
legend style={at={(0.08,-0.2)}, anchor=south west, legend cell align=left, align=left, fill=white}
]

\addplot [color=blue, line width=1.5pt, mark=diamond, mark options={solid, blue}]
  table[row sep=crcr]{%
2	-2.25410E-06\\
4	-2.25119E-06\\
8 -2.25100E-06 \\
16	-2.25073E-06\\
32	-2.24021E-06\\
64	-2.21906E-06\\
};
\addlegendentry{LB}

\addplot [color=red, line width=1.5pt, mark=o, mark options={solid, red}]
  table[row sep=crcr]{%
2	-2.25410E-06\\
4	-2.21774E-06\\
8	-2.21800E-06\\
16	-2.21805E-06\\
32	-2.21807E-06\\
64	-2.21808E-06\\
};
\addlegendentry{GLB}

\addplot [color=mycolor1, line width=1.5pt, mark=triangle, mark options={solid, mycolor1}]
  table[row sep=crcr]{%
2	-2.25100E-06\\
4	-2.25100E-06\\
8	-2.25093E-06\\
16	-2.24728E-06\\
32	-2.22129E-06\\
64	-2.17758E-06\\
};
\addlegendentry{IGA, cubic}

\addplot [color=orange, line width=1.5pt]
  table[row sep=crcr]{%
0.8	-2.21804E-06\\
2	-2.21804E-06\\
4	-2.21804E-06\\
8	-2.21804E-06\\
16	-2.21804E-06\\
32	-2.21804E-06\\
100	-2.21804E-06\\
};
\addlegendentry{$w_A^{\text{ref}}$}

\end{axis}
\end{tikzpicture}%
}
\caption{Convergence of deflection $w_A$ for the rectangular plate \ref{plate} subjected to concentrated load, thickness $h=$ 10$^{-5}$. IGA elements are locked until the elements are refined to a certain number.}
\label{plate_conv}
\end{figure}
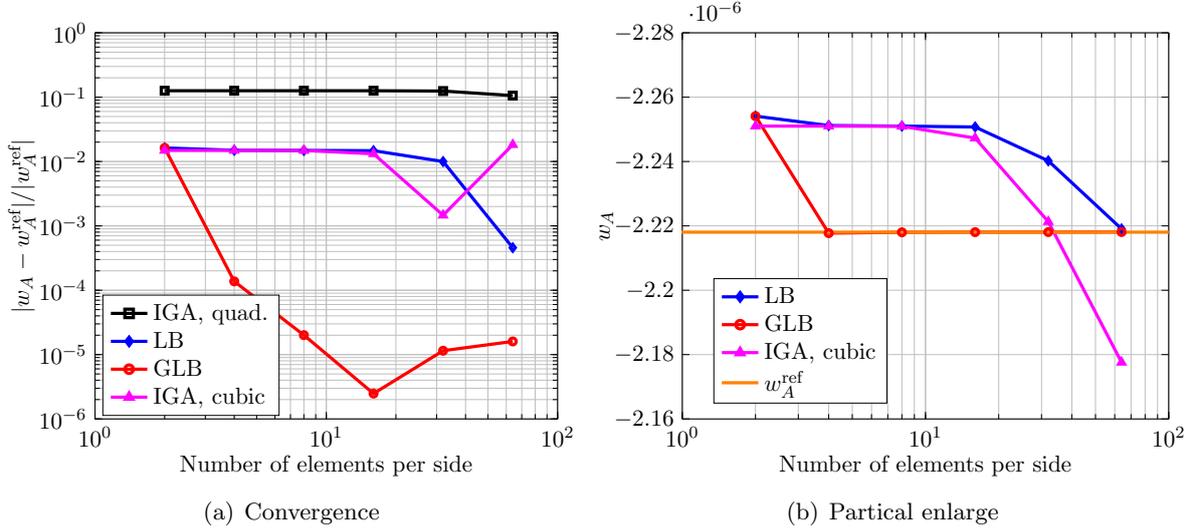

\fref{plate_conv}(b) compares the convergence behavior by a severe locking case
in which the plate thickness is $h=$ 10$^{-5}$.
In this study, elements by IGA of order 2 are locked even by more than 1000 elements, the convergence rate is nearly zero.
Elements by IGA of order 3 start with smaller error, but suffer from locking until the elements are refined to a certain number.
It is inferred that classical IGA elements are locked until the elements are refined to a certain number, and higher order elements could reach this number earlier, this conclusion is already known in literatures.
LB behaves the same as GLB for mesh $2\times 2$.
GLB achieves a good convergence rate at beginning, but in general the convergence rate is smaller as the meshes are refined, at the last two steps the errors become even larger, which is also observed for IGA of order 3.
This could be explained as the results slightly mismatch the reference value,
as pointed out in \fref{plate_conv}(b).

\begin{figure}[htbp]
\centering
\def\svgwidth{0.3 \columnwidth}
\subfigure[Expansion]{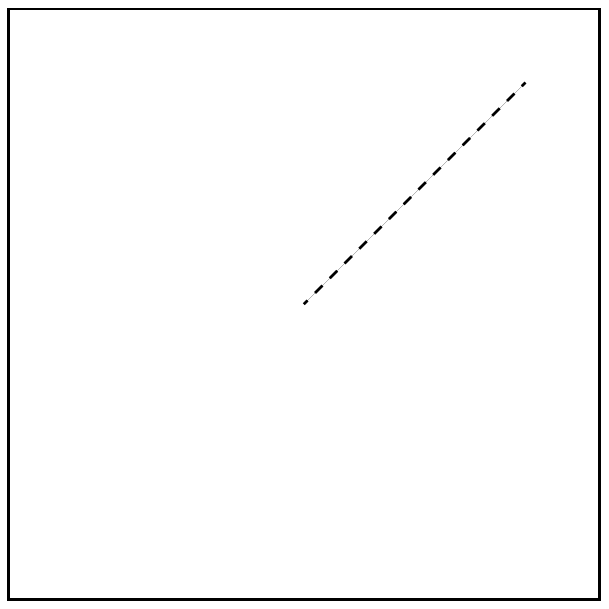}
\def\svgwidth{0.3 \columnwidth}
\subfigure[Rotation]{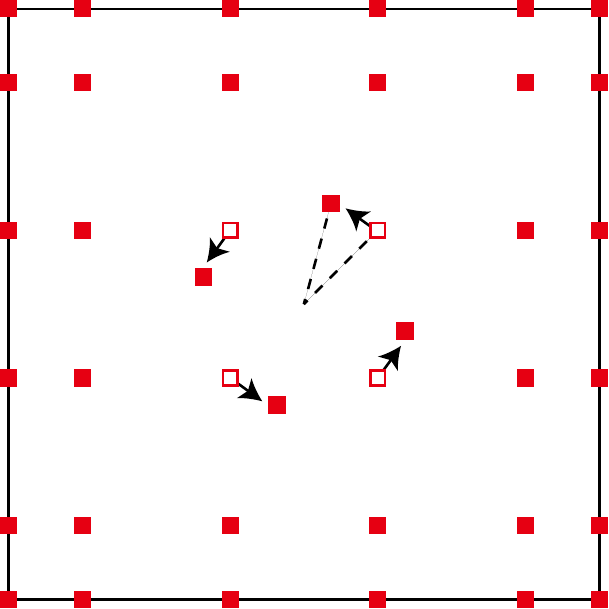}
\caption{Illustration of control mesh distortions. In (a), four selected control points are moved along the diagonal. In (b), four selected control points are moved around the center of the patch. Indexes $e$ and $r$ are employed to indicate the stages of distortions.}
\label{mesh_distortion}
\end{figure}

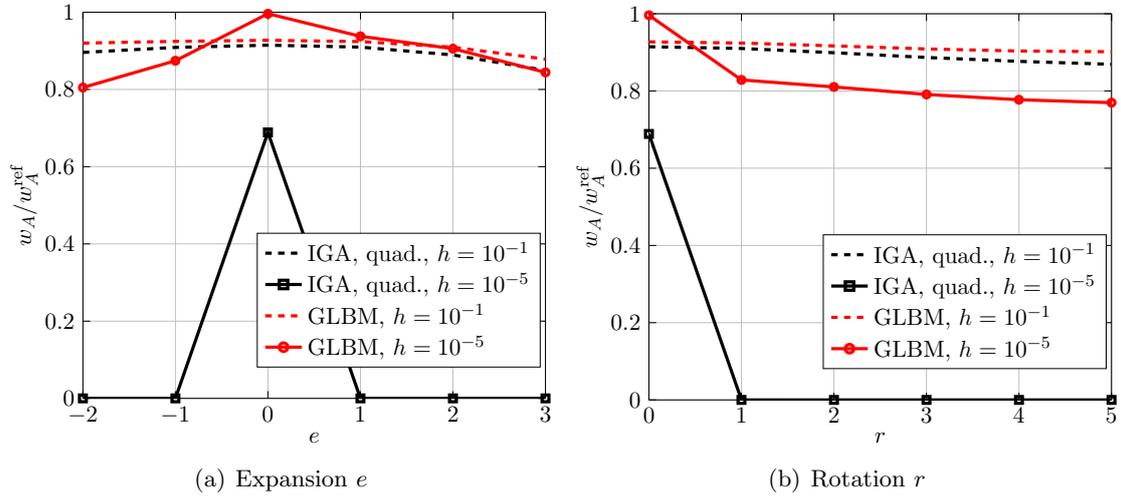
\begin{figure}[htbp]
\centering
\subfigure[Expansion $e$]{% This file was created by matlab2tikz.
\begin{tikzpicture}

\begin{axis}[%
width=0.951\figurewidth,
height=\figureheight,
at={(0\figurewidth,0\figureheight)},
scale only axis,
separate axis lines,
every outer x axis line/.append style={black},
every x tick label/.append style={font=\color{black}},
every x tick/.append style={black},
xmin=-2,
xmax=3,
xtick={-2, -1,  0,  1,  2,  3},
xlabel={$e$},
every outer y axis line/.append style={black},
every y tick label/.append style={font=\color{black}},
every y tick/.append style={black},
ymin=0,
ymax=1,
ylabel={$w_A / w_A^{\text{ref}}$},
axis background/.style={fill=white},
xmajorgrids,
ymajorgrids,
legend style={at={(0.47,0.1)}, anchor=south west, legend cell align=left, align=left, fill=white}
]
\addplot [color=black, dashed, line width=1.5pt]
  table[row sep=crcr]{%
-2	0.89584826597563\\
-1	0.90882514760587\\
0	0.914602687113096\\
1	0.909645147629332\\
2	0.889282400265591\\
3	0.849278364786219\\
};
\addlegendentry{IGA, quad., $h=10^{-1}$}

\addplot [color=black, line width=1.5pt, mark=square, mark options={solid, black}]
  table[row sep=crcr]{%
-2	1.67883891583308e-06\\
-1	6.14753839547338e-06\\
0	0.688747221380216\\
1	9.55600497625543e-06\\
2	3.60049131049813e-06\\
3	1.55172811205416e-06\\
};
\addlegendentry{IGA, quad., $h=10^{-5}$}

\addplot [color=red, dashed, line width=1.5pt]
  table[row sep=crcr]{%
-2	0.920007179399347\\
-1	0.924498996409127\\
0	0.927224118089388\\
1	0.923930040741918\\
2	0.909355390825619\\
3	0.878540282647791\\
};
\addlegendentry{GLBM, $h=10^{-1}$}

\addplot [color=red, line width=1.5pt, mark=o, mark options={solid, red}]
  table[row sep=crcr]{%
-2	0.804942469940386\\
-1	0.874295872486612\\
0	0.996097302212792\\
1	0.937597883196928\\
2	0.905597971607558\\
3	0.844147847832676\\
};
\addlegendentry{GLBM, $h=10^{-5}$}

\end{axis}

\end{tikzpicture}%
}
\subfigure[Rotation $r$]{% This file was created by matlab2tikz.
\begin{tikzpicture}

\begin{axis}[%
width=0.951\figurewidth,
height=\figureheight,
at={(0\figurewidth,0\figureheight)},
scale only axis,
separate axis lines,
every outer x axis line/.append style={black},
every x tick label/.append style={font=\color{black}},
every x tick/.append style={black},
xmin=0,
xmax=5,
xtick={0, 1, 2, 3, 4, 5},
xlabel={$r$},
every outer y axis line/.append style={black},
every y tick label/.append style={font=\color{black}},
every y tick/.append style={black},
ymin=0,
ymax=1,
ylabel={$w_A / w_A^{\text{ref}}$},
axis background/.style={fill=white},
xmajorgrids,
ymajorgrids,
legend style={at={(0.47,0.1)}, anchor=south west, legend cell align=left, align=left, fill=white}
]
\addplot [color=black, dashed, line width=1.5pt]
  table[row sep=crcr]{%
0	0.914602687113096\\
1	0.910087408013946\\
2	0.899086034308613\\
3	0.886782514643277\\
4	0.87672900934847\\
5	0.869126119288301\\
};
\addlegendentry{IGA, quad., $h=10^{-1}$}

\addplot [color=black, line width=1.5pt, mark=square, mark options={solid, black}]
  table[row sep=crcr]{%
0	0.688747221380216\\
1	2.23164216429221e-05\\
2	6.02490022228958e-06\\
3	2.87852379508942e-06\\
4	1.71852897342629e-06\\
5	1.18812992573507e-06\\
};
\addlegendentry{IGA, quad., $h=10^{-5}$}

\addplot [color=red, dashed, line width=1.5pt]
  table[row sep=crcr]{%
0	0.927224118089388\\
1	0.924150584381991\\
2	0.916747122081463\\
3	0.908876764202483\\
4	0.903566120273709\\
5	0.901646921363288\\
};
\addlegendentry{GLBM, $h=10^{-1}$}

\addplot [color=red, line width=1.5pt, mark=o, mark options={solid, red}]
  table[row sep=crcr]{%
0	0.996097302212792\\
1	0.828642833686976\\
2	0.810441295342023\\
3	0.790994114378094\\
4	0.777333093361625\\
5	0.769728200464787\\
};
\addlegendentry{GLBM, $h=10^{-5}$}

\end{axis}
\end{tikzpicture}%
}
\caption{Normalized results of $w_A$ of the rectangular plate \ref{plate} with control mesh distortions \ref{mesh_distortion}. GLB means the generalized local $\bar{B}$. $h$ stands for thickness. When locking and mesh distortion occur at the same time, errors of IGA of order 2 drop down quickly, while GLB keeps good accuracy even for severe distortions.}
\label{distortion_res}
\end{figure}

From the above, it is opined that for plates, when the slenderness ratio reaches a large number, e.g. 10$^3$ or 10$^5$, the classical IGA elements suffer from locking.
However, the studies above are done in an ideal condition that the plates are discreted by structured meshes.
The influence of the mesh distortion on the performance of the proposed formulation is investigated by considering two different kinds of mesh distortions as shown in \fref{mesh_distortion}.
%The plate is assumed to be subjected to concentrated point load at the center of the plate and with simply supported edges.
%Two different plate thickness are considered for this study, viz., $h=$ 10$^{-1}$ and $h=$ 10$^{-5}$.
Thickness $h=$ 10$^{-5}$ is considered here as the case when both locking and mesh distortion appear, and $h=$ 10$^{-1}$ is chosen as the control case when only mesh distortion appears.
The influence of mesh distortion on the normalized center displacement is depicted in \fref{distortion_res}.
When locking and mesh distortion occur at the same time, errors of IGA drop down quickly, while GLB keeps good accuracy even for severe distortions.
It can be inferred that the results with classical IGA deteriorates with mesh distortion, while the proposed formulation is less sensitive to the mesh distortion.

\subsection{Scordelis-Lo roof \label{section42}}

\begin{figure}[htbp]
\centering
\def\svgwidth{0.7\columnwidth}
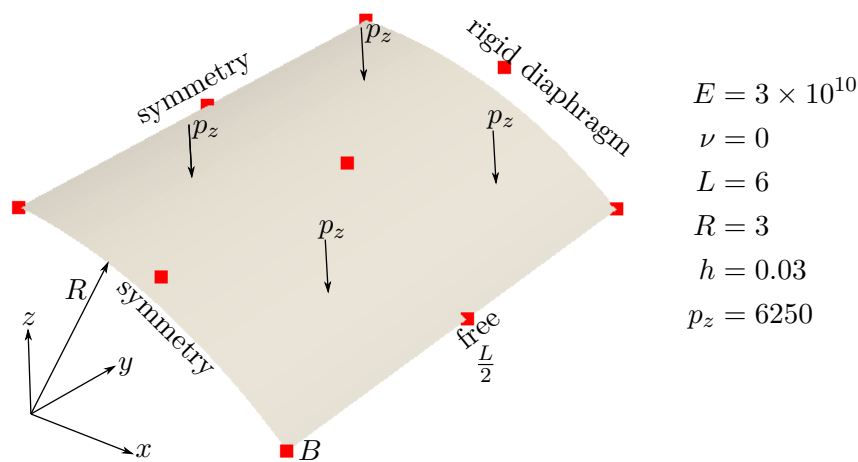
\caption{Scordelis-Lo roof problem in Section \ref{section42}: geometry and boundary conditions. The red filled squares are the corresponding control points. The mid-surface of the cylindrical panel is modeled and the roof is subjected to a uniform pressure.}
\label{roof}
\end{figure}

Next to demonstrate the performance of the proposed formulation when a structure experiences membrane locking, Scordelis-Lo roof problem is considered.
It features a cylindrical panel with ends supported by rigid diaphragm.
The roof is subjected to uniform pressure, $p_z=$ 6250 N/m$^2$ and the vertical displacement of the mid-point of the side edge is monitored to study the convergence behavior.
The material properties are Young's modulus: $E=$ 30 GPa and Poisson's ratio, $\nu=$ 0.
Owing to symmetry only one quarter of the roof is modeled as shown in \fref{roof}. 
The roof is modeled with the control points $(x,y,z)$ and the weights, $w$ given in Table \ref{geo_shell_A}.
The analytical solution based on the deep shell theory is $w_B^{\rm ref} =$ -0.0361 m ~\cite{macneal1985proposed},
however the results obtained by mesh $100\times 100$ of pure IGA of $p=q=4$, $w_B^{\text{ref}} = -0.0361776$, is taken as reference solution.

\begin{table}[htbp]
\scriptsize
\centering
\caption{Control points and weights for the Scordelis-Lo roof problem \ref{roof}.}
\begin{tabular}{ccccccccccc}
\hline\noalign{\smallskip}
& 1 & 2 & 3&4&5&6& 7&8&9 \\
\hline
$x$ & 0 & 1.091910703  &1.928362829  & 0 & 1.091910703  &1.928362829  &0 & 1.091910703  &1.928362829 \\
$y$ & 0 & 0 & 0 & 1.5 &1.5 & 1.5 & 3 & 3  & 3  \\
$z$ & 3 & 3 & 2.298133329 &  3 & 3 & 2.298133329 & 3 & 3 & 2.298133329   \\
$w$ & 1& 0.9396926208& 1& 1& 0.9396926208& 1& 1& 0.9396926208& 1 \\
\hline
\end{tabular}\label{geo_shell_A}
\end{table}

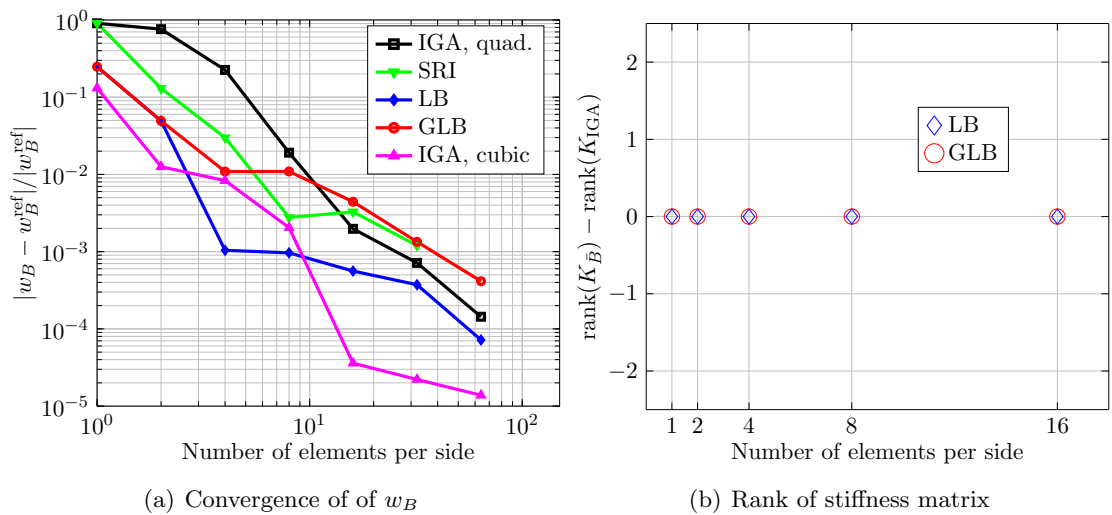
\begin{figure}[htbp]
\centering
\subfigure[Convergence of of $w_B$]{% This file was created by matlab2tikz.
\definecolor{mycolor1}{rgb}{1.00000,0.00000,1.00000}%
\begin{tikzpicture}

\begin{axis}[%
width=0.951\figurewidth,
height=\figureheight,
at={(0\figurewidth,0\figureheight)},
scale only axis,
separate axis lines,
every outer x axis line/.append style={black},
every x tick label/.append style={font=\color{black}},
every x tick/.append style={black},
xmode=log,
xmin=1,
xmax=150,
xminorticks=true,
xlabel={Number of elements per side},
every outer y axis line/.append style={black},
every y tick label/.append style={font=\color{black}},
every y tick/.append style={black},
ymode=log,
ymin=1e-05,
ymax=1,
yminorticks=true,
ylabel={$|w_B-w_B^{\text{ref}}|/|w_B^{\text{ref}}|$},
axis background/.style={fill=white},
xmajorgrids,
xminorgrids,
ymajorgrids,
yminorgrids,
legend style={at={(0.73,0.5)}, anchor=south west, legend cell align=left, align=left, fill=white}
]
\addplot [color=black, line width=1.5pt, mark=square, mark options={solid, black}]
  table[row sep=crcr]{%
1	0.90553159966388\\
2	0.760240038034585\\
4	0.2255207642298\\
8	0.0191278581221529\\
16	0.00197636106319919\\
32	0.000715912608907041\\
64	0.000143735350050774\\
};
\addlegendentry{IGA, quad.}

\addplot [color=green, line width=1.5pt, mark=triangle, mark options={solid, rotate=180, green}]
  table[row sep=crcr]{%
1	0.906793153774711\\
2	0.129986510990226\\
4	0.0299245942240501\\
8	0.00277519791252047\\
16	0.00327274335500433\\
32	0.00119963734465542\\
};
\addlegendentry{SRI}

\addplot [color=blue, line width=1.5pt, mark=diamond, mark options={solid, blue}]
  table[row sep=crcr]{%
1	0.248092742470479\\
2	0.049342687187652\\
4	0.00104208128786859\\
8	0.000964685330149163\\
16	0.000561120693467614\\
32	0.0003731590818628\\
64	7.18676750253869e-05\\
};
\addlegendentry{LB}

\addplot [color=red, line width=1.5pt, mark=o, mark options={solid, red}]
  table[row sep=crcr]{%
1	0.248092742470479\\
2	0.049342687187652\\
4	0.0108851886250055\\
8	0.0109266507452125\\
16	0.00442815443810552\\
32	0.00134060855335898\\
64	0.00041462120206982\\
};
\addlegendentry{GLB}

\addplot [color=mycolor1, line width=1.5pt, mark=triangle, mark options={solid, mycolor1}]
  table[row sep=crcr]{%
1	0.131302242271461\\
2	0.0125878996948387\\
4	0.00831453717217294\\
8	0.00204822873822477\\
16	3.59338375127893e-05\\
32	2.21131307769241e-05\\
64	1.38207067356735e-05\\
};
\addlegendentry{IGA, cubic}

\end{axis}
\end{tikzpicture}%
}
\subfigure[Rank of stiffness matrix]{% This file was created by matlab2tikz.
\begin{tikzpicture}

\begin{axis}[%
width=0.951\figurewidth,
height=\figureheight,
at={(0\figurewidth,0\figureheight)},
scale only axis,
separate axis lines,
every outer x axis line/.append style={black},
every x tick label/.append style={font=\color{black}},
every x tick/.append style={black},
xmin=0,
xmax=18,
xtick={1,2,4,8,16},
xlabel={Number of elements per side},
every outer y axis line/.append style={black},
every y tick label/.append style={font=\color{black}},
every y tick/.append style={black},
ymin=-2.5,
ymax=2.5,
ytick={-2, -1,  0,  1,  2},
ylabel={$\text{rank}(K_{\bar{B}})-\text{rank}(K_{\text{IGA}})$},
axis background/.style={fill=white},
xmajorgrids,
ymajorgrids,
legend style={legend cell align=left, align=left, draw=black}
]

\addplot [only marks, blue, mark=diamond, mark size=4.5pt]
  table[row sep=crcr]{%
1	0\\
2	0\\
4	0\\
8	0\\
16	0\\
};
\addlegendentry{LB}

\addplot [only marks, red, mark=o, mark size=4.5pt]
  table[row sep=crcr]{%
1	0\\
2	0\\
4	0\\
8	0\\
16	0\\
};
\addlegendentry{GLB}

\end{axis}
\end{tikzpicture}%
}
\caption{Results of the Scordelis-Lo roof \ref{roof}. IGA suffers from locking. LB and GLB finally converge.}
\label{result_roof}
\end{figure}

\begin{figure}[htbp]
\centering
\subfigure[$4\times 4$ IGA]{\includegraphics[width=0.4\textwidth]{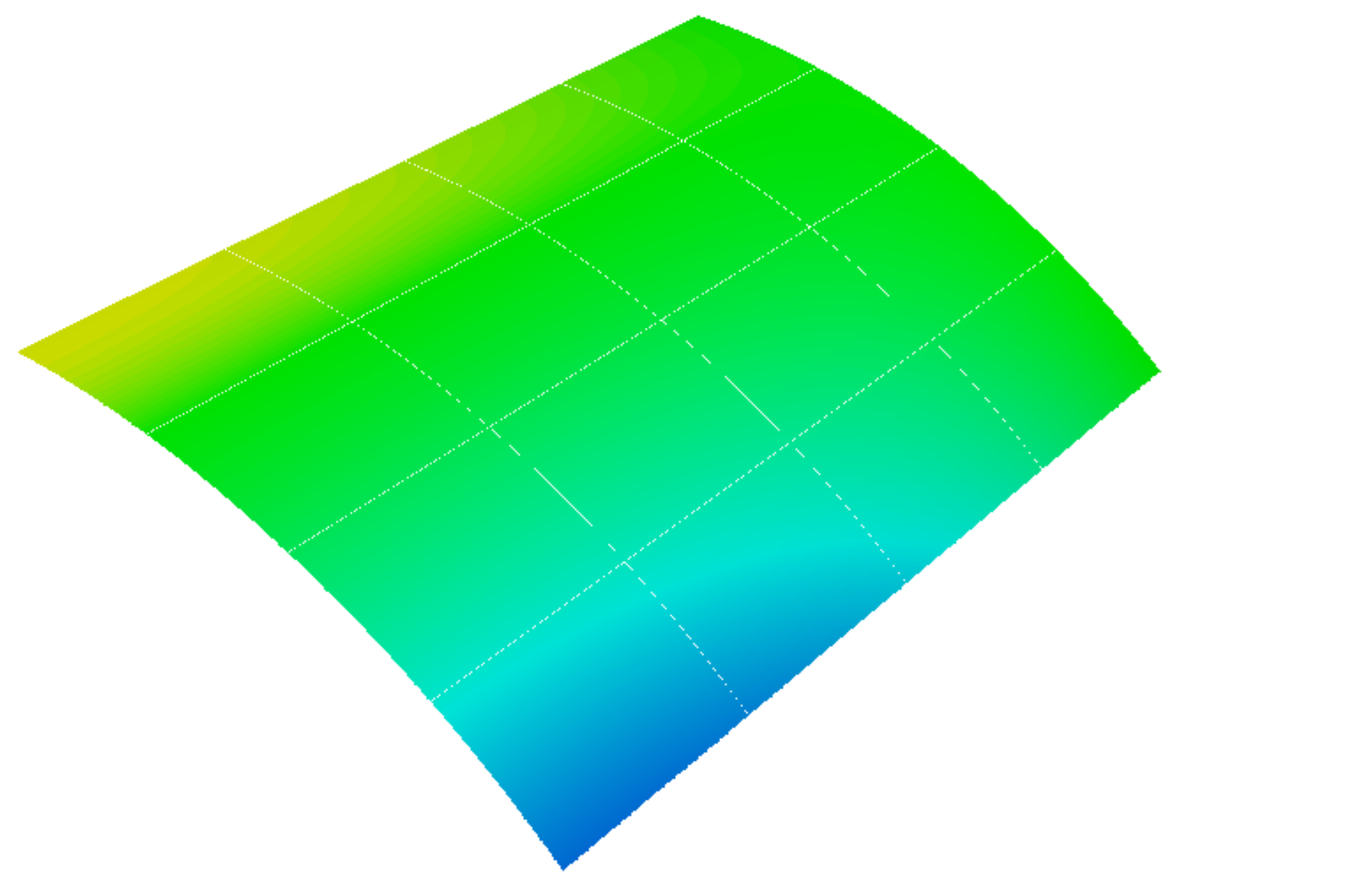}}
\subfigure[$8\times 8$ IGA]{\includegraphics[width=0.4\textwidth]{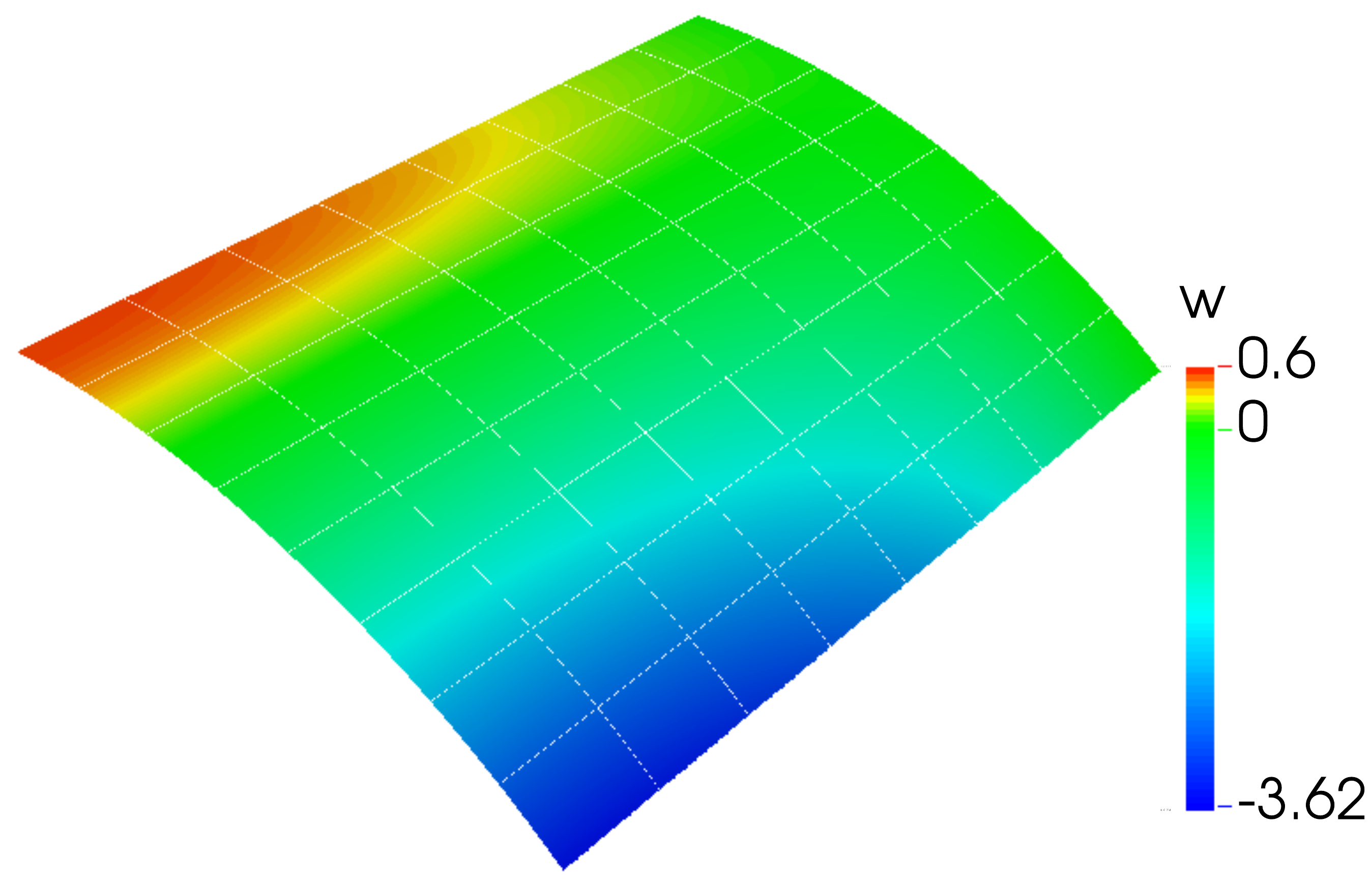}}
\subfigure[$4\times 4$ GLB]{\includegraphics[width=0.4\textwidth]{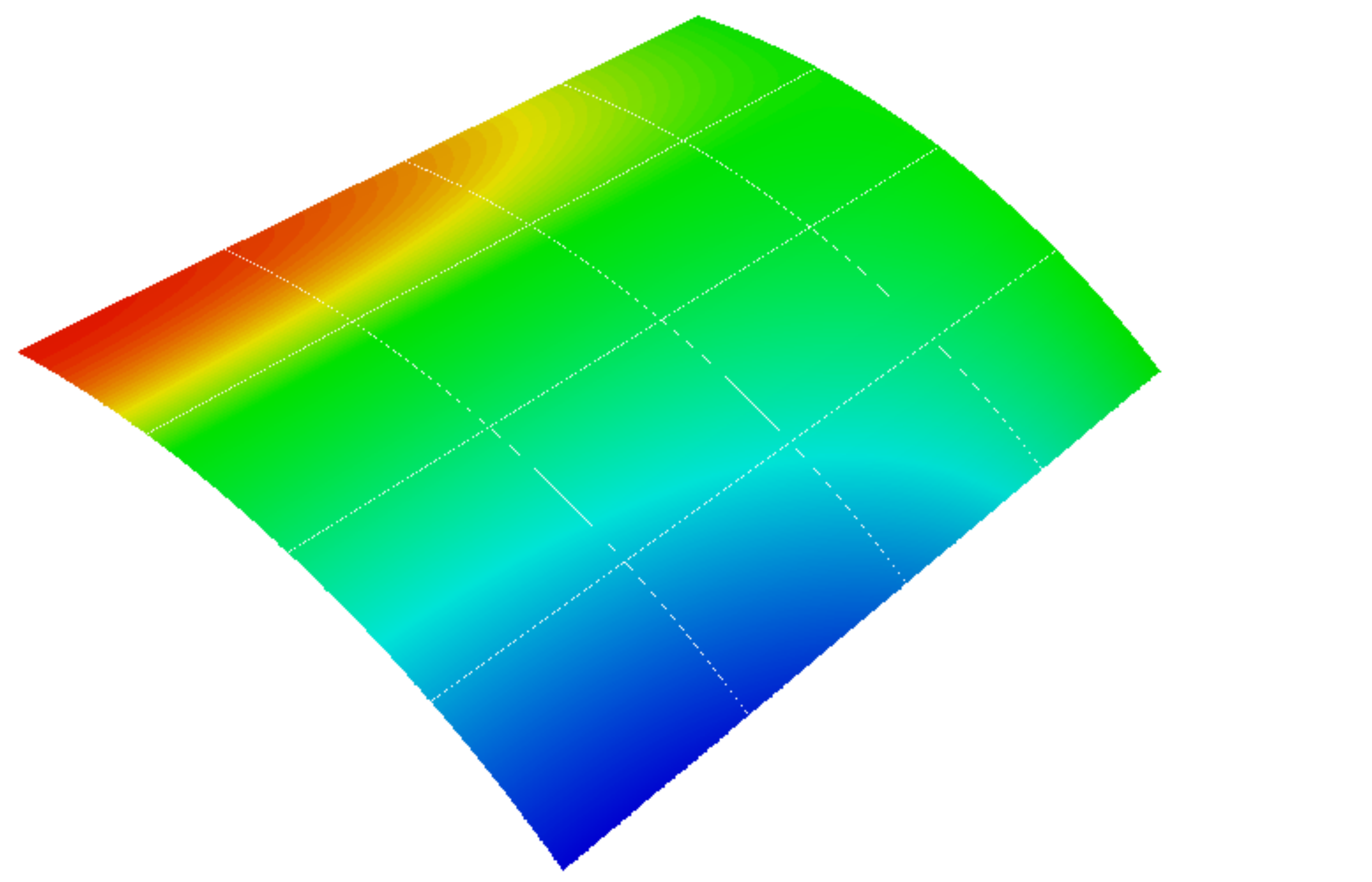}}
\subfigure[$8\times 8$ GLB]{\includegraphics[width=0.4\textwidth]{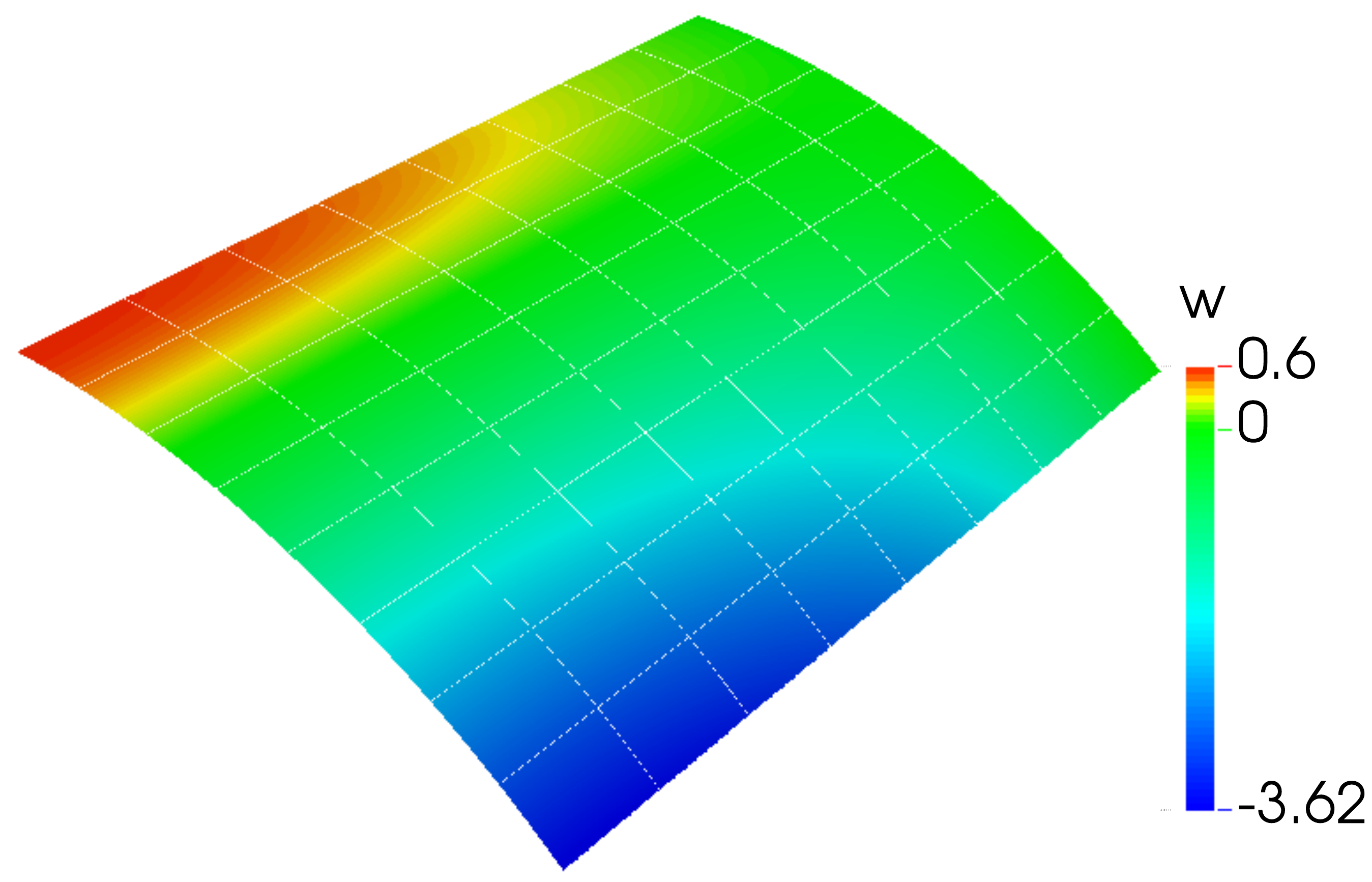}}
\caption{Deflection field $w \times 10^2$ of the Scordelis-Lo roof \ref{roof} by IGA of order 2 (a,b), and by the generalized local $\bar{B}$ method (c,d). For coarse meshes pure IGA is locked, but deformation is captured very well by GLB.}
\label{result_A_cloud}
\end{figure}

For the geometry considered here, $R/h=$ 100 and $L/h=$ 200, the structure experiences membrane locking as the transverse shear strain is negligible.
The roof is dominated by membrane and bending deformations. 
The convergence of the normalized vertical displacement with mesh refinement is shown in \fref{result_roof}.
The results from the proposed formulation is compared with selective reduced integration technique~\cite{adam2015improved}.
It can be seen that except IGA of order 2, the convergence lines by others stall in the middle and then converge as before.
%while GLB captures the stall first and then converge as before.
It is noticed that for coarse meshes the present method leads to rank sufficient matrices,
however for other problems slightly rank deficiency is possible as in the strain smoothing method by a single subcell \cite{bordas2010strain}.
In addition, the contour plot of the deflection $w_B$ by IGA and GLB are given in \fref{result_A_cloud} as a function of mesh refinement.
It is obvious that IGA is locked in the case of coarse mesh, while GLB captures the deformation quite  very well even for coarse meshes.

\subsection{Pinched cylinder \label{section43}}
\begin{figure}[htbp]
\centering
\def\svgwidth{0.7\columnwidth}
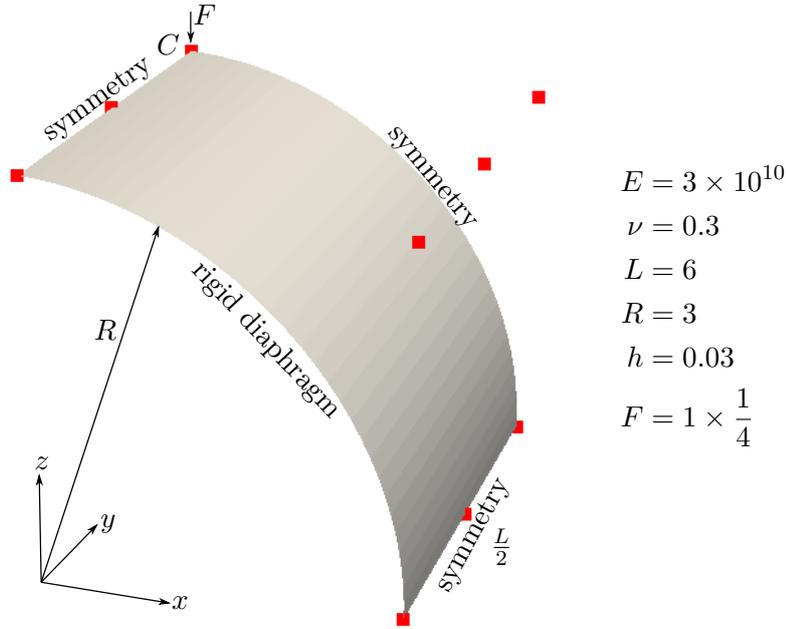
\caption{The mid-surface of a fourth of the pinched cylinder in Section \ref{section43}. The red filled squares are the corresponding control points.}
\label{pin_cyl}
\end{figure}

\begin{figure}[htbp]
\centering
\subfigure[Convergence of of $w_C$]{% This file was created by matlab2tikz.
\definecolor{mycolor1}{rgb}{1.00000,0.00000,1.00000}%
\begin{tikzpicture}

\begin{axis}[%
width=0.951\figurewidth,
height=\figureheight,
at={(0\figurewidth,0\figureheight)},
scale only axis,
separate axis lines,
every outer x axis line/.append style={black},
every x tick label/.append style={font=\color{black}},
every x tick/.append style={black},
xmode=log,
xmin=1,
xmax=150,
xminorticks=true,
xlabel={Number of elements per side},
every outer y axis line/.append style={black},
every y tick label/.append style={font=\color{black}},
every y tick/.append style={black},
ymode=log,
ymin=0.001,
ymax=1,
yminorticks=true,
ylabel={$|w_C-w_C^{\text{ref}}|/|w_C^{\text{ref}}|$},
axis background/.style={fill=white},
xmajorgrids,
xminorgrids,
ymajorgrids,
yminorgrids,
legend style={at={(0.73,0.5)}, anchor=south west, legend cell align=left, align=left, fill=white}
]
\addplot [color=black, line width=1.5pt, mark=square, mark options={solid, black}]
  table[row sep=crcr]{%
1	0.977126953566166\\
2	0.958609835324994\\
4	0.902529821127018\\
8	0.631554463219714\\
16	0.197513203041809\\
32	0.0291219842746663\\
64	0.00588893310817365\\
};
\addlegendentry{IGA, quad.}

\addplot [color=green, line width=1.5pt, mark=triangle, mark options={solid, rotate=180, green}]
  table[row sep=crcr]{%
1	0.978487915586581\\
2	0.95482462273182\\
4	0.623538522765163\\
8	0.111550913725786\\
16	0.00515214421701391\\
32	0.0121435716513751\\
};
\addlegendentry{SRI}

\addplot [color=blue, line width=1.5pt, mark=diamond, mark options={solid, blue}]
  table[row sep=crcr]{%
1	0.89405190865969\\
2	0.953418001312237\\
4	0.86338535672414\\
8	0.460154241645244\\
16	0.0669509847156641\\
32	0.00997085112561988\\
64	0.00368394445579809\\
};
\addlegendentry{LB}

\addplot [color=red, line width=1.5pt, mark=o, mark options={solid, red}]
  table[row sep=crcr]{%
1	0.89405190865969\\
2	0.953418001312237\\
4	0.643847006055652\\
8	0.0862473244345011\\
16	0.00823912833033966\\
32	0.00293639952243178\\
64	0.00225339084230566\\
};
\addlegendentry{GLB}

\addplot [color=mycolor1, line width=1.5pt, mark=triangle, mark options={solid, mycolor1}]
  table[row sep=crcr]{%
1	0.957816899893515\\
2	0.92848630218025\\
4	0.68645384044487\\
8	0.175587011003431\\
16	0.0169676565810844\\
32	0.00555011777866219\\
64	0.00232868313775271\\
};
\addlegendentry{IGA, cubic}

\end{axis}
\end{tikzpicture}%
}
\subfigure[Rank of stiffness matrix]{% This file was created by matlab2tikz.
\begin{tikzpicture}

\begin{axis}[%
width=0.951\figurewidth,
height=\figureheight,
at={(0\figurewidth,0\figureheight)},
scale only axis,
separate axis lines,
every outer x axis line/.append style={black},
every x tick label/.append style={font=\color{black}},
every x tick/.append style={black},
xmin=0,
xmax=18,
xtick={1,2,4,8,16},
xlabel={Number of elements per side},
every outer y axis line/.append style={black},
every y tick label/.append style={font=\color{black}},
every y tick/.append style={black},
ymin=-2.5,
ymax=2.5,
ytick={-2, -1,  0,  1,  2},
ylabel={$\text{rank}(K_{\bar{B}})-\text{rank}(K_{\text{IGA}})$},
axis background/.style={fill=white},
xmajorgrids,
ymajorgrids,
legend style={legend cell align=left, align=left, draw=black}
]

\addplot [only marks, blue, mark=diamond, mark size=4.5pt]
  table[row sep=crcr]{%
1	0\\
2	0\\
4	0\\
8	0\\
16	0\\
};
\addlegendentry{LB}

\addplot [only marks, red, mark=o, mark size=4.5pt]
  table[row sep=crcr]{%
1	0\\
2	0\\
4	0\\
8	0\\
16	0\\
};
\addlegendentry{GLB}

\end{axis}
\end{tikzpicture}%
}
\caption{Results of the pinched cylinder \ref{pin_cyl}. GLB achieves a good accuracy and convergence.}
\label{pin_cyl_res}
\end{figure}
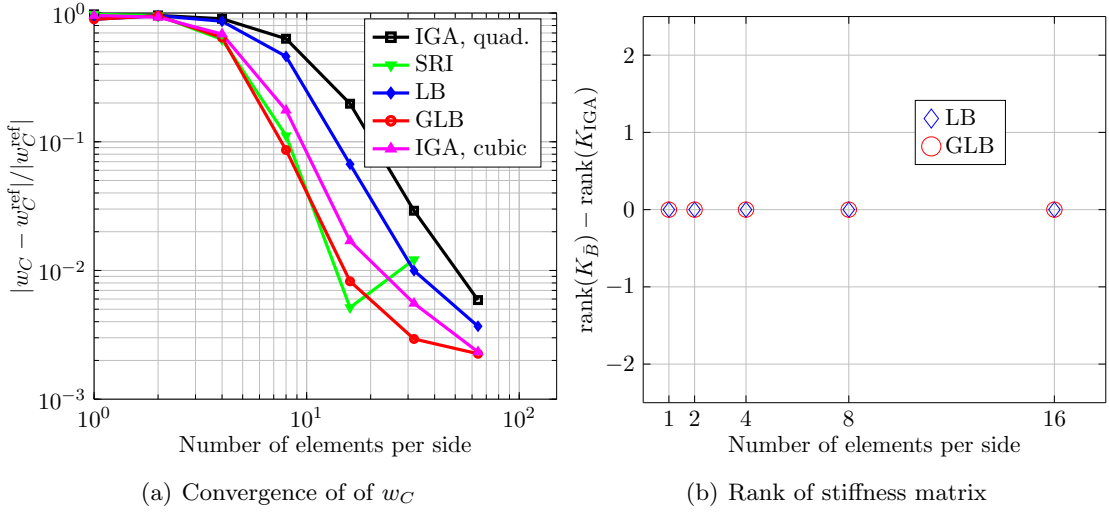

\begin{figure}[htbp]
\centering
\subfigure[$4\times 4$ IGA]{\includegraphics[width=0.4\textwidth]{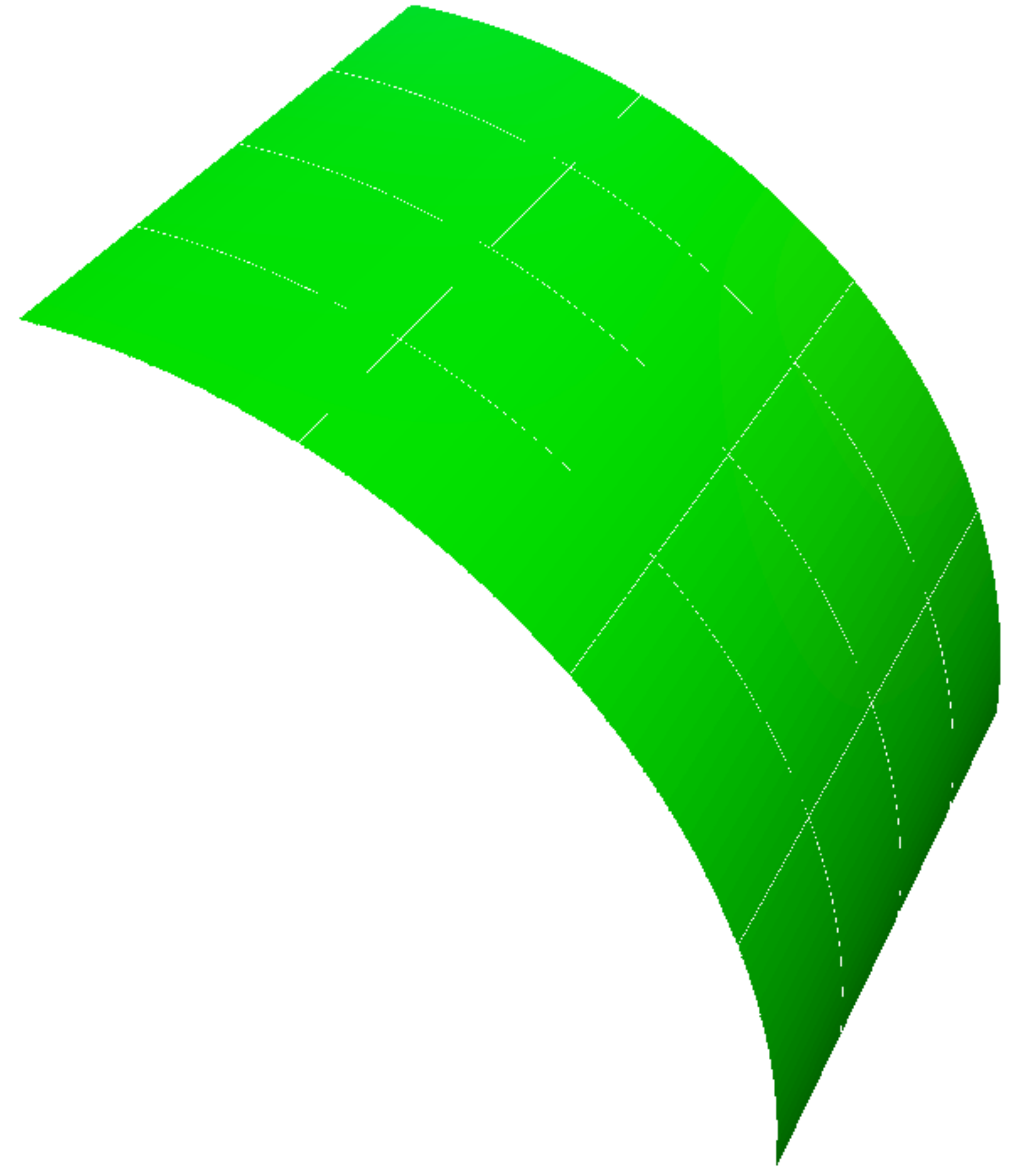}}
\subfigure[$8\times 8$ IGA]{\includegraphics[width=0.4\textwidth]{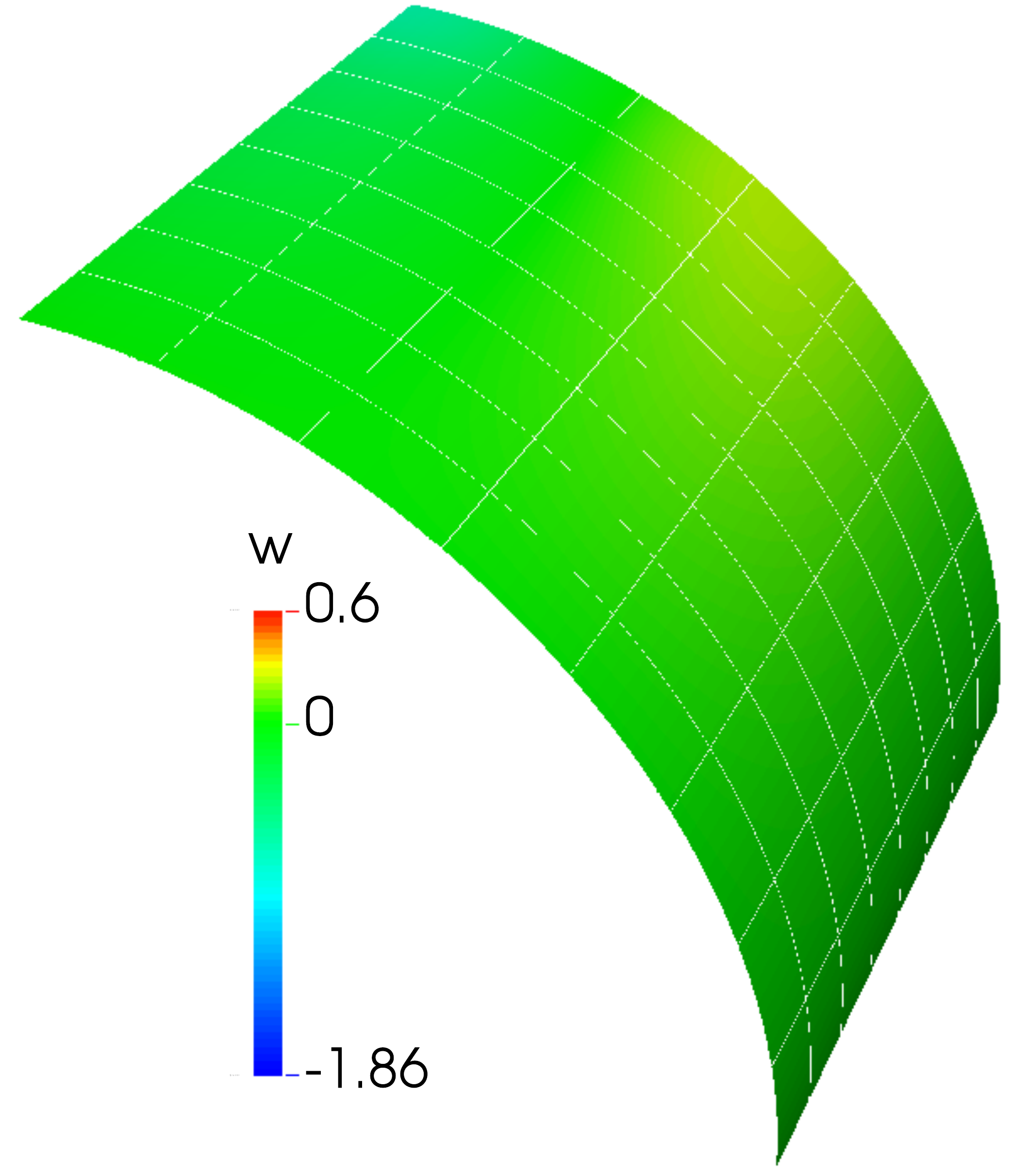}}
\subfigure[$4\times 4$ GLB]{\includegraphics[width=0.4\textwidth]{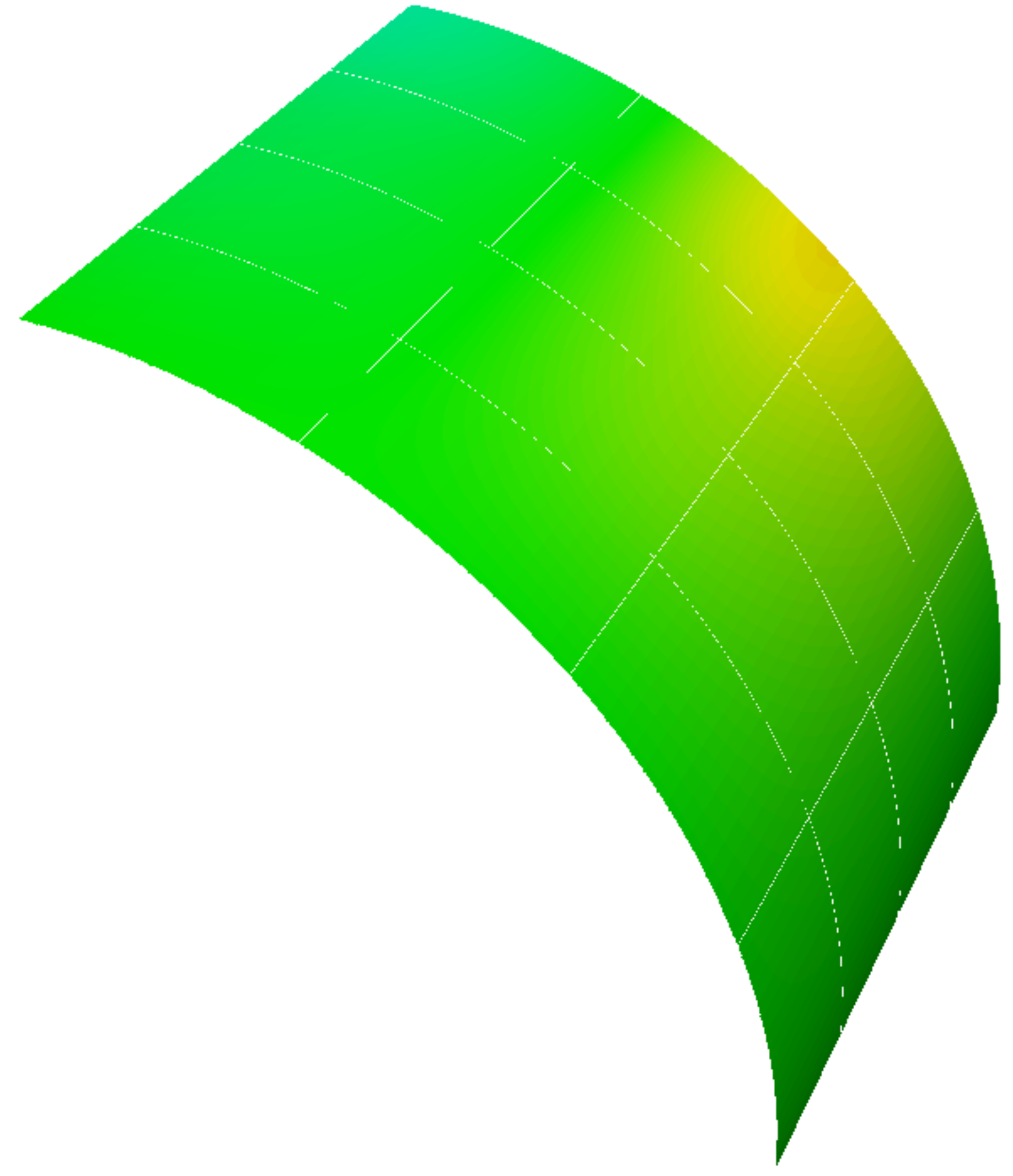}}
\subfigure[$8\times 8$ GLB]{\includegraphics[width=0.4\textwidth]{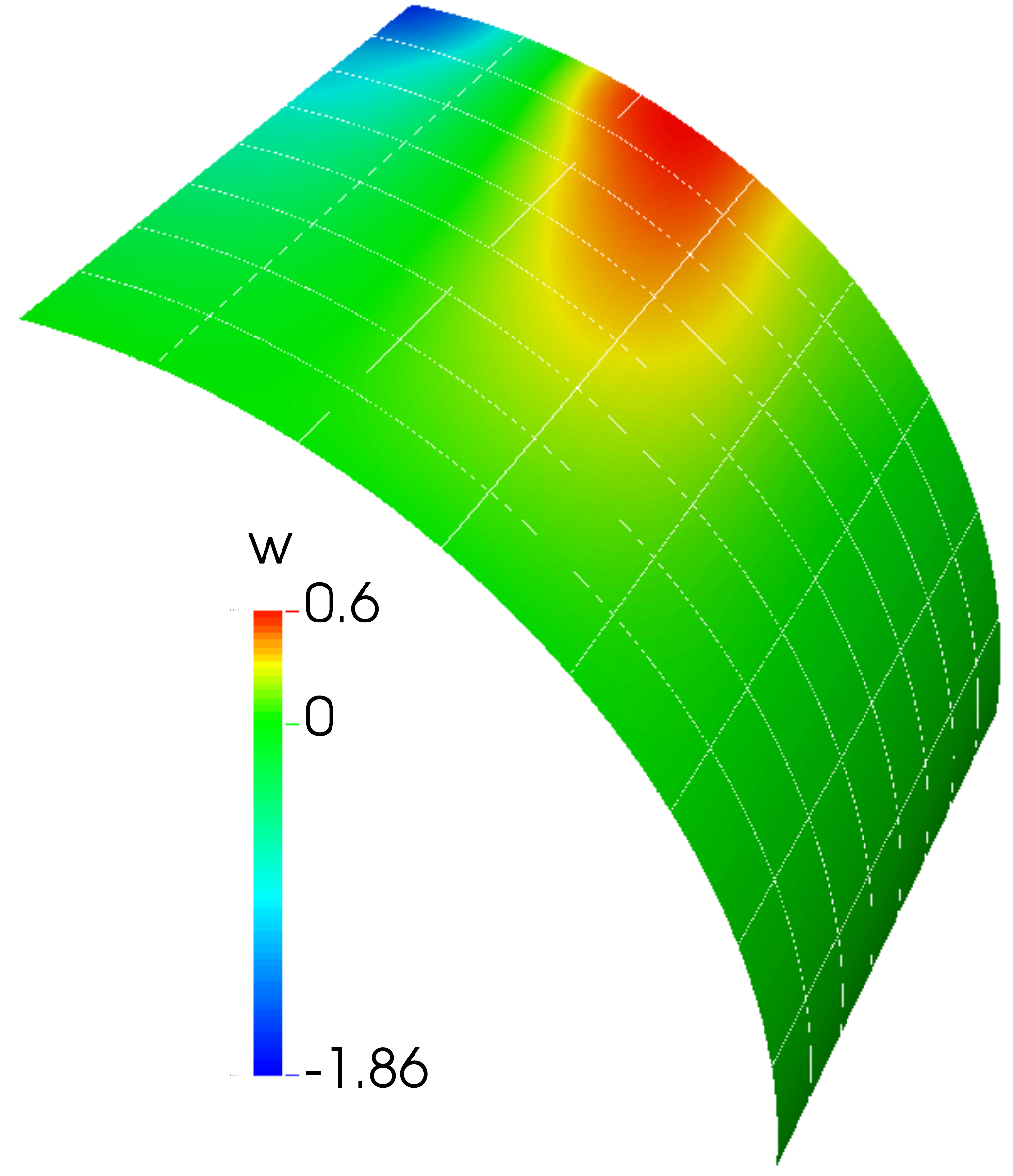}}
\caption{Field of $w \times 10^7$ of the pinched cylinder \ref{pin_cyl} by IGA of order 2 (a,b), and by the generalized local $\bar{B}$ method (c,d). IGA is locked for coarse meshes and refined meshes. For GLB the elements seem more flexible than IGA to be deformed.}
\label{result_B_cloud}
\end{figure}

From the above two examples, it is clear that the proposed formulation yields accurate results when the structure experience either shear locking or membrane locking.
To demonstrate the robustness of the proposed formulation when the structure experiences both shear and membrane locking, we consider the pinched cylinder problem.
Again, due to symmetry only one quarter of the cylinder is modeled as shown in \fref{pin_cyl}.
The corresponding control points and weights are given in Table \ref{geo_shell_B}.
The cylinder is made up of homogeneous isotropic material with Young's modulus, $E=$ 30 GPa and Poisson's ratio, $\nu=$ 0.3.
The concentrated load acting on the cylinder is $P=$ 0.25 N.
The reference value of the vertical displacement is take as $w_C^{\text{ref}}=$ -1.85942$^{-7}$ m.,
which is obtained by mesh $100\times 100$ of pure IGA of order 4.
This example serves a test case to evaluate the performance when the structure is dominated by bending behaviour.
The convergence of the vertical displacement with mesh refinement is shown in \fref{pin_cyl_res} and it is evident that the proposed formulation yields more accurate results than the conventional IGA of order 2,
in addition the $\bar{B}$ formulation does not change the rank of the stiffness matrix as in  \fref{pin_cyl_res}(b).
The contour plot of $w_C$ is shown in \fref{result_B_cloud}, the elements by GLB seem more flexible than IGA to be deformed.

\begin{table}[htbp]
%\scriptsize
\centering
\caption{Control points and weights for the pinched cylinder problem \ref{pin_cyl}.}
\begin{tabular}{ccccccccccc}
\hline\noalign{\smallskip}
& 1 & 2 & 3&4&5&6& 7&8&9 \\
\hline$x$ & 0 & 3  &3  & 0 & 3  &3  &0 & 3  &3 \\
$y$ & 0 & 0 & 0 & 1.5 &1.5 & 1.5 & 3 & 3  & 3  \\
$z$ & 3 & 3 & 0 &  3 & 3 & 0 & 3 & 3 & 0   \\
$w$ & 1& 0.7071067812& 1& 1& 0.7071067812& 1& 1& 0.7071067812& 1 \\
\hline
	\end{tabular}\label{geo_shell_B}
\end{table}

\subsection{Pinched hemisphere with hole \label{section44}}
\begin{figure}[htbp]
\centering
\def\svgwidth{0.7\columnwidth}
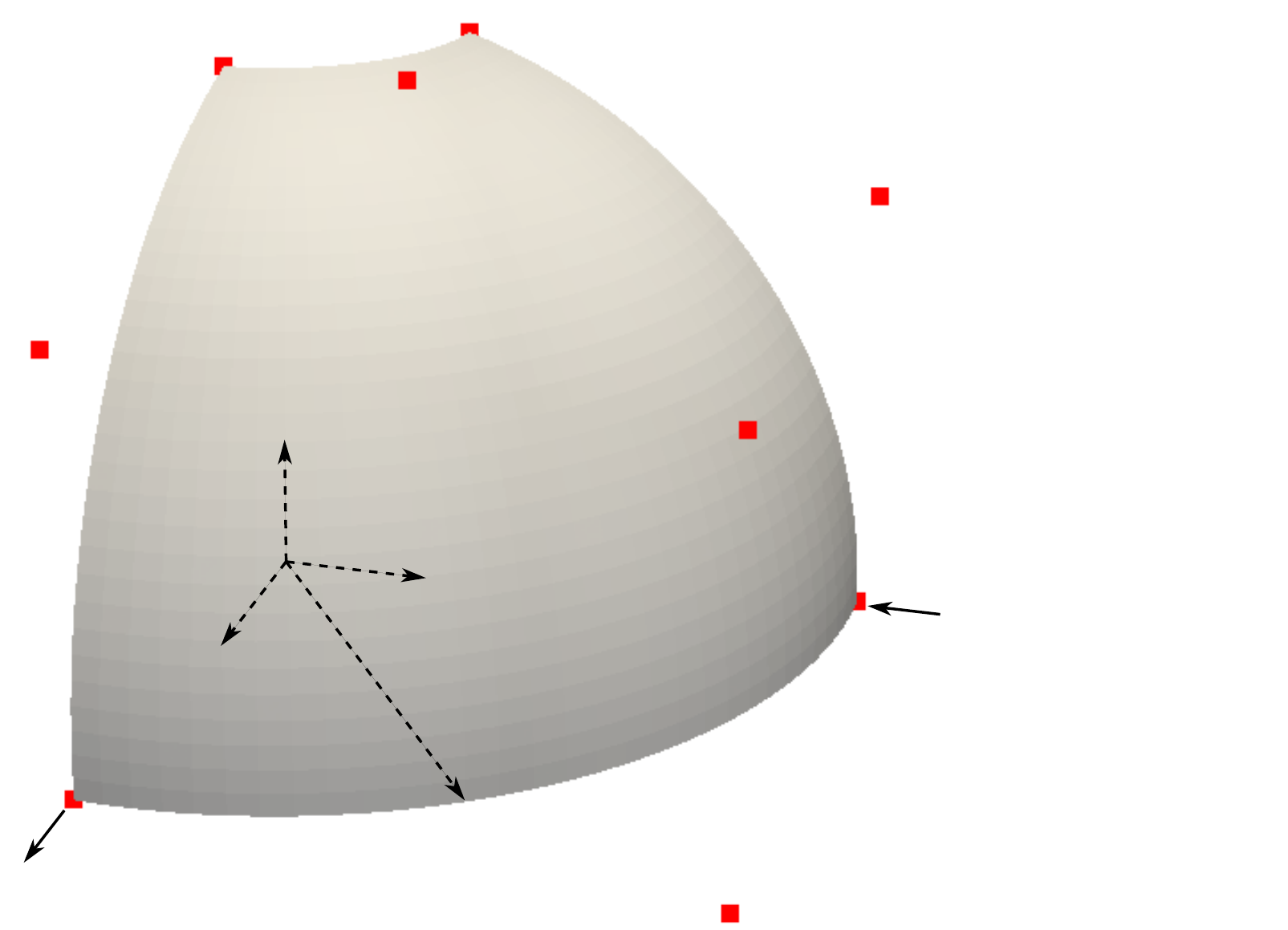
\caption{The mid-surface of a fourth of the pinched hemisphere in Section \ref{section44}. The red filled squares are the corresponding control points.}
\label{pin_hem}
\end{figure}

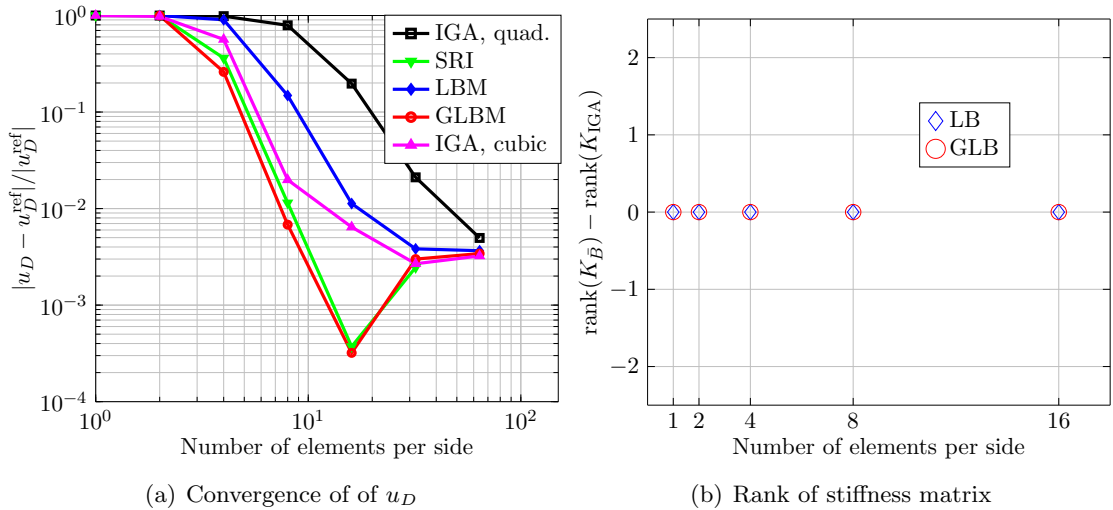
\begin{figure}[htbp]
\centering
\subfigure[Convergence of of $u_D$]{% This file was created by matlab2tikz.
\definecolor{mycolor1}{rgb}{1.00000,0.00000,1.00000}%
\begin{tikzpicture}

\begin{axis}[%
width=0.951\figurewidth,
height=\figureheight,
at={(0\figurewidth,0\figureheight)},
scale only axis,
separate axis lines,
every outer x axis line/.append style={black},
every x tick label/.append style={font=\color{black}},
every x tick/.append style={black},
xmode=log,
xmin=1,
xmax=150,
xminorticks=true,
xlabel={Number of elements per side},
every outer y axis line/.append style={black},
every y tick label/.append style={font=\color{black}},
every y tick/.append style={black},
ymode=log,
ymin=0.0001,
ymax=1,
yminorticks=true,
ylabel={$|u_D-u_D^{\text{ref}}|/|u_D^{\text{ref}}|$},
axis background/.style={fill=white},
xmajorgrids,
xminorgrids,
ymajorgrids,
yminorgrids,
legend style={at={(0.78,0.52)}, anchor=south west, legend cell align=left, align=left, fill=white}
]
\addplot [color=black, line width=1.5pt, mark=square, mark options={solid, black}]
  table[row sep=crcr]{%
1	0.999835395744681\\
2	0.99903334893617\\
4	0.984685957446809\\
8	0.791287234042553\\
16	0.196976595744681\\
32	0.0210702127659574\\
64	0.00496276595744675\\
};
\addlegendentry{IGA, quad.}

\addplot [color=green, line width=1.5pt, mark=triangle, mark options={solid, rotate=180, green}]
  table[row sep=crcr]{%
1	0.999840425531915\\
2	0.997914893617021\\
4	0.362457446808511\\
8	0.0114574468085106\\
16	0.000372340425531844\\
32	0.00245744680851059\\
};
\addlegendentry{SRI}

\addplot [color=blue, line width=1.5pt, mark=diamond, mark options={solid, blue}]
  table[row sep=crcr]{%
1	1.0553085106383\\
2	0.997885180851064\\
4	0.903493936170213\\
8	0.148373404255319\\
16	0.0112702127659575\\
32	0.0038265957446809\\
64	0.0036617021276596\\
};
\addlegendentry{LBM}

\addplot [color=red, line width=1.5pt, mark=o, mark options={solid, red}]
  table[row sep=crcr]{%
1	1.0553085106383\\
2	0.997885180851064\\
4	0.260557446808511\\
8	0.0068127659574468\\
16	0.000320212765957501\\
32	0.00299787234042552\\
64	0.00341808510638294\\
};
\addlegendentry{GLBM}

\addplot [color=mycolor1, line width=1.5pt, mark=triangle, mark options={solid, mycolor1}]
  table[row sep=crcr]{%
1	0.99904090212766\\
2	0.978358617021277\\
4	0.565168085106383\\
8	0.0198574468085106\\
16	0.00643085106382972\\
32	0.00268191489361708\\
64	0.00323085106382983\\
};
\addlegendentry{IGA, cubic}

\end{axis}
\end{tikzpicture}%
}
\subfigure[Rank of stiffness matrix]{% This file was created by matlab2tikz.
\begin{tikzpicture}

\begin{axis}[%
width=0.951\figurewidth,
height=\figureheight,
at={(0\figurewidth,0\figureheight)},
scale only axis,
separate axis lines,
every outer x axis line/.append style={black},
every x tick label/.append style={font=\color{black}},
every x tick/.append style={black},
xmin=0,
xmax=18,
xtick={1,2,4,8,16},
xlabel={Number of elements per side},
every outer y axis line/.append style={black},
every y tick label/.append style={font=\color{black}},
every y tick/.append style={black},
ymin=-2.5,
ymax=2.5,
ytick={-2, -1,  0,  1,  2},
ylabel={$\text{rank}(K_{\bar{B}})-\text{rank}(K_{\text{IGA}})$},
axis background/.style={fill=white},
xmajorgrids,
ymajorgrids,
legend style={legend cell align=left, align=left, draw=black}
]

\addplot [only marks, blue, mark=diamond, mark size=4.5pt]
  table[row sep=crcr]{%
1	0\\
2	0\\
4	0\\
8	0\\
16	0\\
};
\addlegendentry{LB}

\addplot [only marks, red, mark=o, mark size=4.5pt]
  table[row sep=crcr]{%
1	0\\
2	0\\
4	0\\
8	0\\
16	0\\
};
\addlegendentry{GLB}

\end{axis}
\end{tikzpicture}%
}
\caption{Results of $u_D$ of the pinched hemisphere \ref{pin_hem}.}
\label{pin_hem_res}
\end{figure}

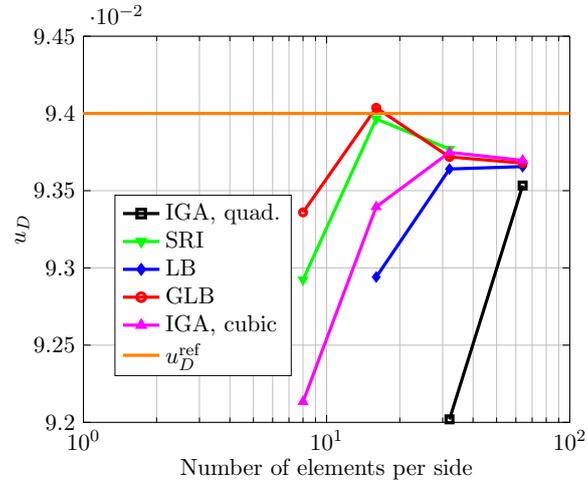
\begin{figure}[htbp]
\centering
\definecolor{mycolor1}{rgb}{1.00000,0.00000,1.00000}%
\begin{tikzpicture}

\begin{axis}[%
width=\figurewidth,
height=\figureheight,
at={(0\figurewidth,0\figureheight)},
scale only axis,
unbounded coords=jump,
separate axis lines,
every outer x axis line/.append style={black},
every x tick label/.append style={font=\color{black}},
every x tick/.append style={black},
xmode=log,
xmin=1,
xmax=100,
xminorticks=true,
xlabel={Number of elements per side},
every outer y axis line/.append style={black},
every y tick label/.append style={font=\color{black}},
every y tick/.append style={black},
ymin=0.092,
ymax=0.0945,
ylabel={$u_D$},
axis background/.style={fill=white},
xmajorgrids,
xminorgrids,
ymajorgrids,
legend style={at={(0.08,0.15)}, anchor=south west, legend cell align=left, align=left, fill=white}
]

\addplot [color=black, line width=1.5pt, mark=square, mark options={solid, black}]
  table[row sep=crcr]{%
%1	1.54728e-05\\
%2	9.08652e-05\\
%4	0.00143952\\
%8	0.019619\\
%16	0.0754842\\
32	0.0920194\\
64	0.0935335\\
};
\addlegendentry{IGA, quad.}

\addplot [color=green, line width=1.5pt, mark=triangle, mark options={solid, rotate=180, green}]
  table[row sep=crcr]{%
%1	1.5e-05\\
%2	0.000196\\
%4	0.059929\\
%7.7	 0.091 \\
8	0.092923\\
16	0.093965\\
32	0.093769\\
64	nan\\
};
\addlegendentry{SRI}

\addplot [color=blue, line width=1.5pt, mark=diamond, mark options={solid, blue}]
  table[row sep=crcr]{%
%1	0.11\\
%2	0.000198793\\
%4	0.00907157\\
%8	0.0800529\\
%15  0.091 \\
16	0.0929406\\
32	0.0936403\\
64	0.0936558\\
};
\addlegendentry{LB}

\addplot [color=red, line width=1.5pt, mark=o, mark options={solid, red}]
  table[row sep=crcr]{%
%1	0.11\\
%2	0.000198793\\
%4	0.0695076\\
%1.2 0.0946 \\
%1.3 0.091 \\
%7.5  0.091 \\
8	0.0933596\\
16	0.0940367\\
32	0.0937182\\
64	0.0936787\\
};
\addlegendentry{GLB}

\addplot [color=mycolor1, line width=1.5pt, mark=triangle, mark options={solid, mycolor1}]
  table[row sep=crcr]{%
%1	9.01552e-05\\
%2	0.00203429\\
%4	0.0408742\\
%7.9  0.091 \\
8	0.0921334\\
16	0.0933955\\
32	0.0937479\\
64	0.0936963\\
};
\addlegendentry{IGA, cubic}

\addplot [color=orange, line width=1.5pt]
  table[row sep=crcr]{%
0.8	0.094\\
2	0.094\\
4	0.094\\
8	0.094\\
16	0.094\\
32	0.094\\
100	0.094\\
};
\addlegendentry{$u_D^{\text{ref}}$}

\end{axis}
\end{tikzpicture}%
\caption{Partial enlarged of results $u_D$, converged results using various methods all mismatch the reference value.}
\label{pin_hem_res_enlarged}
\end{figure}

\begin{figure}[htbp]
\centering
\subfigure[$4\times 4$ IGA]{\includegraphics[width=0.4\textwidth]{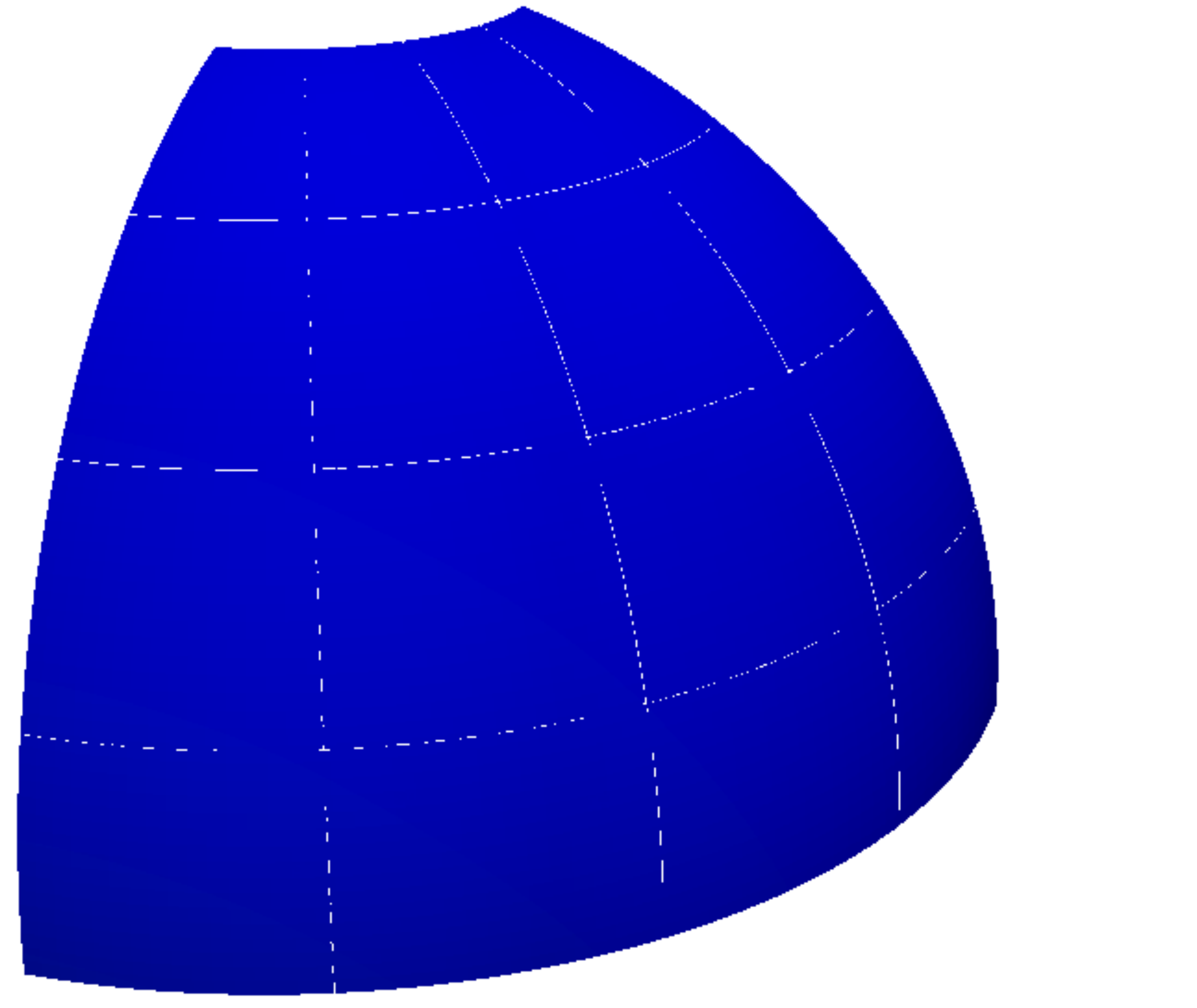}}
\subfigure[$8\times 8$ IGA]{\includegraphics[width=0.4\textwidth]{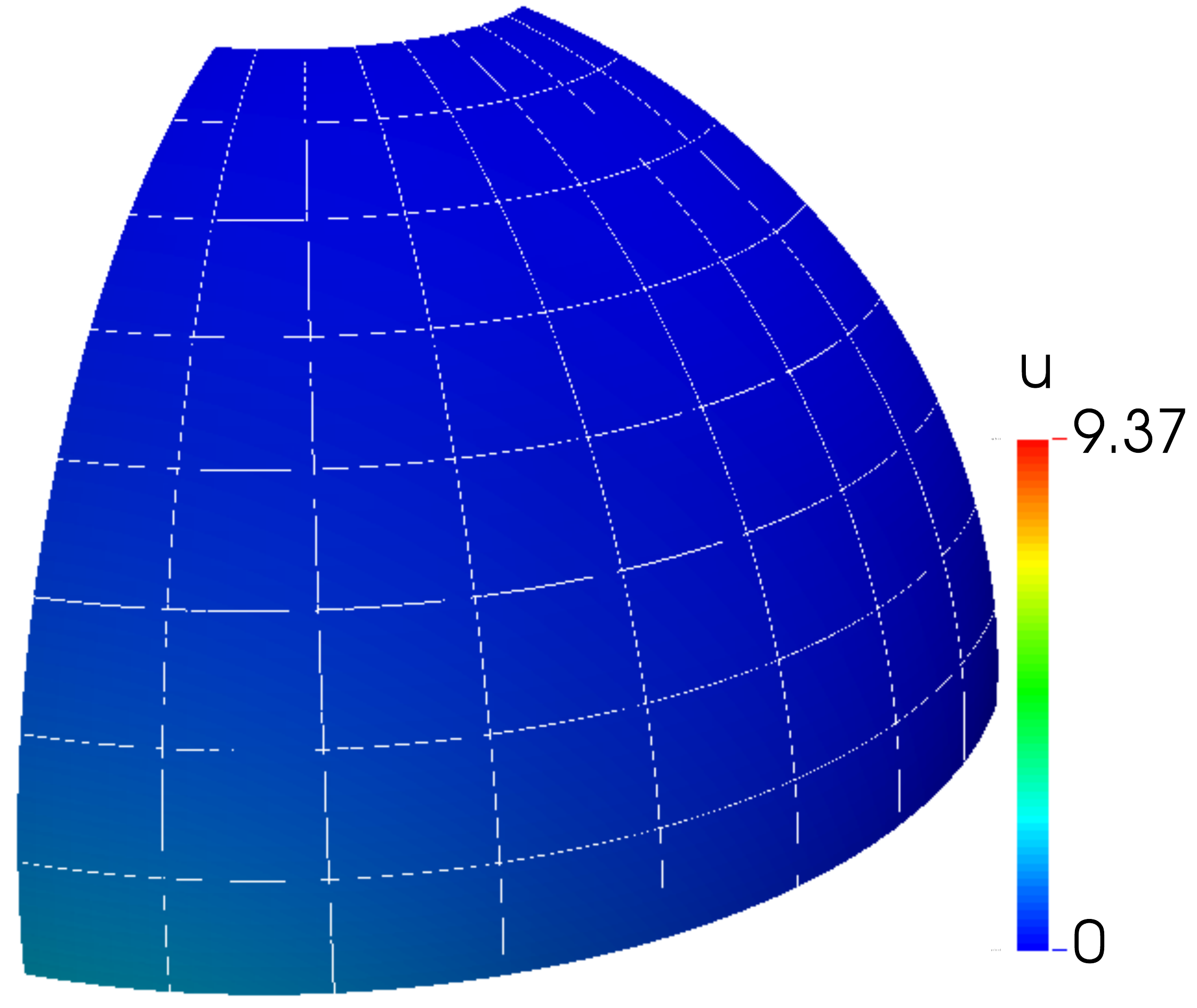}}
\subfigure[$4\times 4$ GLB]{\includegraphics[width=0.4\textwidth]{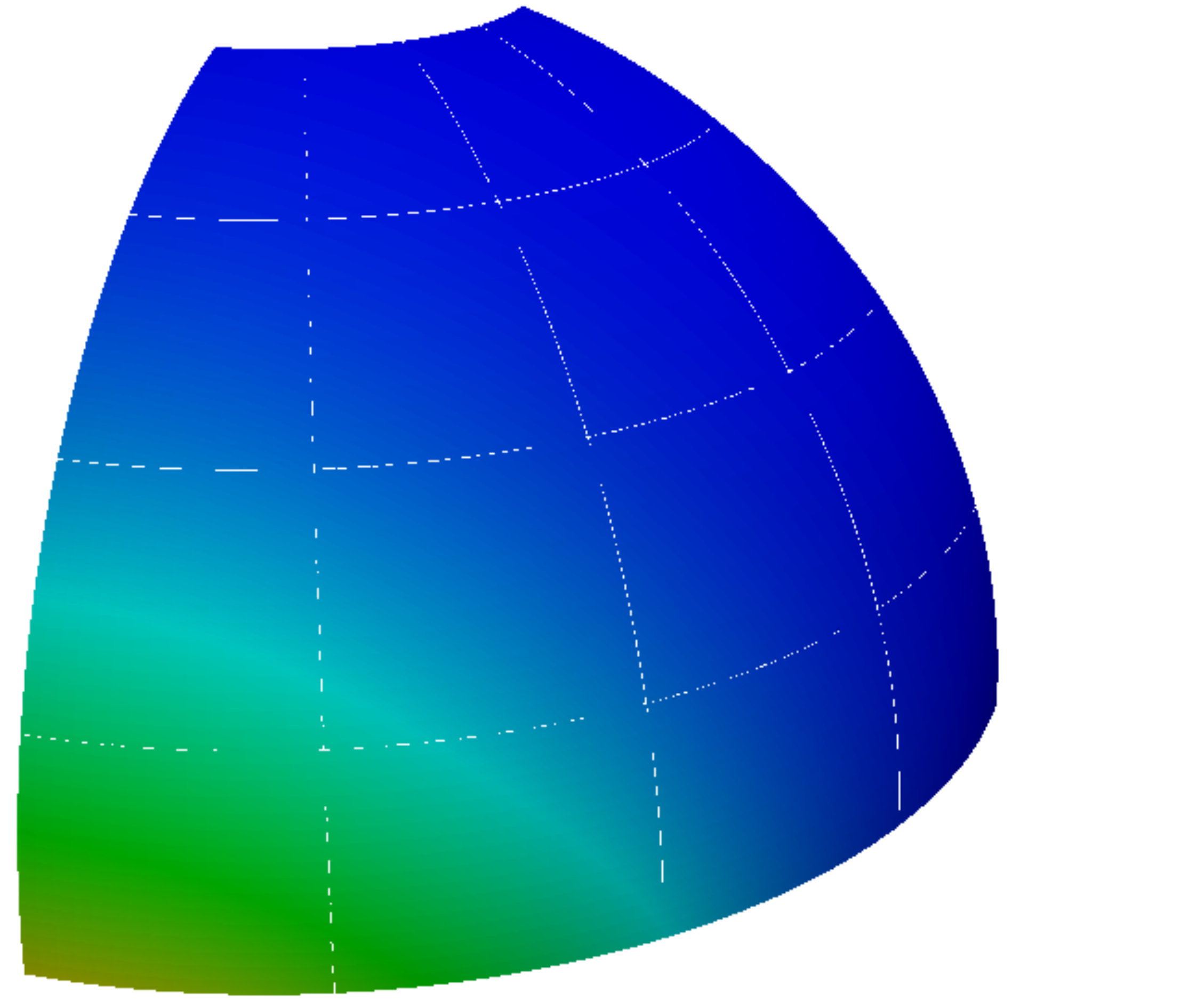}}
\subfigure[$8\times 8$ GLB]{\includegraphics[width=0.4\textwidth]{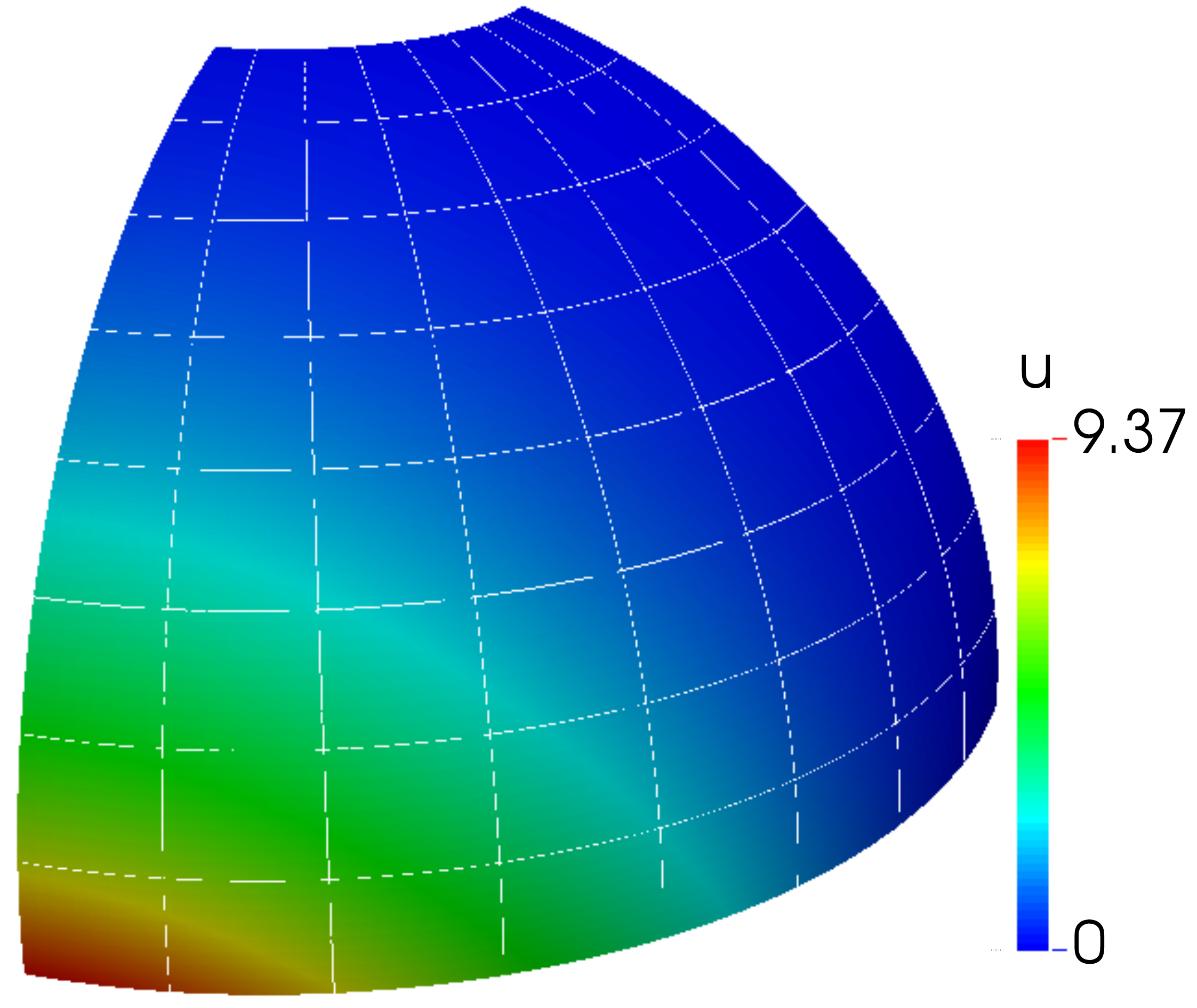}}
\caption{Field of $u \times 10^2$ of the pinched hemisphere \ref{pin_hem} by IGA of order 2 (a,b), and by the generalized local $\bar{B}$ method (c,d). Severe locking is noticed by pure IGA, while elements deform more easily for GLB.}
\label{result_C_cloud}
\end{figure}

As the last example, consider a pinched hemisphere with 18$^\circ$ hole subjected to equal and opposite concentrated forces applied at the four cardinal points.
The hemisphere is modeled with Young's modulus, $E=$ 68.25 MPa, Poisson's ratio $\nu=$ 0.3 and concentrated force, $P=$ 1 N.
As before, owing to symmetry, only one quadrant of the hemisphere is modeled as shown in \fref{pin_hem}. The location of the control points is also shown.
The control points and weights employed to model the hemisphere are given in Table \ref{geo_shell_C}.
This example experiences severe membrane and shear locking, has right body rotations and the discretization experiences severe mesh distortion.
The mesh distortion further enhances the locking pathology, as shown in the plate example.
To evaluate the convergence properties, the horizontal displacement $u_D^{\text{ref}}=$ 0.0940 m \cite{adam2015improved} is taken as the reference solution.
The results form the proposed formulation are plotted in \fref{pin_hem_res}.
%in which numerical instability of $\bar{B}$ methods (LB and GLB) is observed for the initial single element case.
As the elements are refined, GLB behaves similarly to SRI, but the results obtained by GLB is slightly accurate.
Once again the rank of the stiffness is not changed by $\bar{B}$ method.
The reason that prevents the error to go below is the mismatch between the convergence value and the adopted reference value, as shown in \fref{pin_hem_res_enlarged}.
Moreover, the field of displacement $u_x$ is plotted in \fref{result_C_cloud}, it seems that GLB has more abilities to capture the deformations.

\begin{table}[htbp]
\scriptsize
\centering
\caption{Control points and the corresponding weights for the pinched hemisphere problem \ref{pin_hem}.}
\begin{tabular}{ccccccccccc}
\hline\noalign{\smallskip}
& 1 & 2 & 3&4&5&6& 7&8&9 \\
$x$ & 10 & 10 & 0 & 10 & 10 & 0 & 3.090169944  & 3.090169944  & 0 \\
$y$ & 0 & 10 & 10 & 0 &10 & 10 & 0 & 3.090169944  & 3.090169944  \\
$z$ & 0 & 0 & 0 & 7.265425281 & 7.265425281 & 7.265425281 & 9.510565163 & 9.510565163  & 9.510565163  \\
$w$ & 1 &  0.7071067810 &  1 &  0.8090169942 &  0.5720614025 &  0.8090169942 &  1 &  0.7071067810  & 1 \\
\hline
\end{tabular}\label{geo_shell_C}
\end{table}

\section{Conclusion remarks}
The local $\bar{B}$ method is adopted to unlock the degenerated Reissner-Mindlin plate and shell elements within the framework of isogeometric analysis.
The plate/shell mid-surface and the unknown field is described with non-uniform rational B-splines.
The proposed method uses multiple sets of lower order B-spline basis functions as projection bases, by which the locking strains are modified in the sense of $L_2$ projection, in this way field-consistent strains are obtained.

The salient features of the proposed local $\bar{B}$ method are:
(a) has less computational effort than classical $\bar{B}$;
(b) yields better accuracy than classical IGA especially in cases of coarse meshes and mesh distortions;
(c) suppresses both shear and membrane locking commonly encountered when lower order elements are employed and suitable for both thick and thin models;

Future work includes extending the approach to large deformations, large deflections and large rotations as well as investigating the behaviour of the stabilization technique for enriched approximations such as those encountered in partition of unity methods \cite{bordas2010strain,chen2012extended,surendran2017linear}.

\section*{Acknowledgments}
Q. Hu is funded by China Scholarship Council and National Natural Science Foundation of China (No. 11272075).
Y. Xia is funded by National Natural Science Foundation of China (No.61572021, 11272075).
St\'ephane Bordas thanks partial funding for his time provided by the European Research Council Starting Independent Research Grant (ERC Stg grant agreement No. 279578) "RealTCut Towards real time multiscale simulation of cutting in non-linear materials with applications to surgical simulation and computer guided surgery".
We also thank the funding from the Luxembourg National Research Fund (INTER/MOBILITY/14/8813215/CBM/Bordas and INTER/FWO/15/10318764).
Q. Hu is thankful for Prof. Gengdong Cheng for the valuable suggestions of this research subject.

\section*{References}
\bibliographystyle{elsarticle-num}
\bibliography{GLB}

\begin{thebibliography}{10}
\expandafter\ifx\csname url\endcsname\relax
  \def\url#1{\texttt{#1}}\fi
\expandafter\ifx\csname urlprefix\endcsname\relax\def\urlprefix{URL }\fi
\expandafter\ifx\csname href\endcsname\relax
  \def\href#1#2{#2} \def\path#1{#1}\fi

\bibitem{hughes2005isogeometric}
T.~J. Hughes, J.~A. Cottrell, Y.~Bazilevs, Isogeometric analysis: {CAD}, finite
  elements, {NURBS}, exact geometry and mesh refinement, Computer methods in
  applied mechanics and engineering 194~(39) (2005) 4135--4195.

\bibitem{kagan1998new}
P.~Kagan, A.~Fischer, P.~Z. Bar-Yoseph, New {B}-spline finite element approach
  for geometrical design and mechanical analysis, International Journal for
  Numerical Methods in Engineering 41~(3) (1998) 435--458.

\bibitem{lipton2010robustness}
S.~Lipton, J.~A. Evans, Y.~Bazilevs, T.~Elguedj, T.~J. Hughes, Robustness of
  isogeometric structural discretizations under severe mesh distortion,
  Computer Methods in Applied Mechanics and Engineering 199~(5) (2010)
  357--373.

\bibitem{marussig2015fast}
B.~Marussig, J.~Zechner, G.~Beer, T.-P. Fries, Fast isogeometric boundary
  element method based on independent field approximation, Computer Methods in
  Applied Mechanics and Engineering 284 (2015) 458--488.

\bibitem{xu2014geometry}
G.~Xu, E.~Atroshchenko, S.~Bordas, Geometry-independent field approximation for
  spline-based finite element methods, in: Proceedings of the 11th World
  Congress in Computational Mechanics, 2014.

\bibitem{xu2014geometry2}
G.~Xu, E.~Atroshchenko, W.~Ma, S.~Bordas, Geometry-{I}ndependent {F}ield
  approxima{T}ion: {CAD}-analysis integration, geometrical exactness and
  adaptivity, Computer Methods in Applied Mechanics \& Engineering.

\bibitem{atroshchenko2017weakening}
E.~Atroshchenko, G.~Xu, S.~Tomar, S.~Bordas, {Weakening the tight coupling
  between geometry and simulation in isogeometric analysis: from sub-and
  super-geometric analysis to Geometry Independent Field approximaTion (GIFT)},
  arXiv preprint arXiv:1706.06371.

\bibitem{toshniwal2017smooth}
D.~Toshniwal, H.~Speleers, T.~J. Hughes, Smooth cubic spline spaces on
  unstructured quadrilateral meshes with particular emphasis on extraordinary
  points: {G}eometric design and isogeometric analysis considerations, Computer
  Methods in Applied Mechanics and Engineering.

\bibitem{kiendl2009isogeometric}
J.~Kiendl, K.-U. Bletzinger, J.~Linhard, R.~W{\"u}chner, Isogeometric shell
  analysis with {K}irchhoff--{L}ove elements, Computer Methods in Applied
  Mechanics and Engineering 198~(49) (2009) 3902--3914.

\bibitem{benson2011large}
D.~Benson, Y.~Bazilevs, M.-C. Hsu, T.~Hughes, A large deformation,
  rotation-free, isogeometric shell, Computer Methods in Applied Mechanics and
  Engineering 200~(13) (2011) 1367--1378.

\bibitem{benson2010isogeometric}
D.~Benson, Y.~Bazilevs, M.-C. Hsu, T.~Hughes, {Isogeometric shell analysis: the
  Reissner--Mindlin shell}, Computer Methods in Applied Mechanics and
  Engineering 199~(5) (2010) 276--289.

\bibitem{benson2013blended}
D.~Benson, S.~Hartmann, Y.~Bazilevs, M.-C. Hsu, T.~Hughes, Blended isogeometric
  shells, Computer Methods in Applied Mechanics and Engineering 255 (2013)
  133--146.

\bibitem{dornisch2013isogeometric}
W.~Dornisch, S.~Klinkel, B.~Simeon, Isogeometric {R}eissner--{M}indlin shell
  analysis with exactly calculated director vectors, Computer Methods in
  Applied Mechanics and Engineering 253 (2013) 491--504.

\bibitem{hosseini2013isogeometric}
S.~Hosseini, J.~J. Remmers, C.~V. Verhoosel, R.~Borst, An isogeometric
  solid-like shell element for nonlinear analysis, International Journal for
  Numerical Methods in Engineering 95~(3) (2013) 238--256.

\bibitem{cottrell2006isogeometric}
J.~A. Cottrell, A.~Reali, Y.~Bazilevs, T.~J. Hughes, Isogeometric analysis of
  structural vibrations, Computer methods in applied mechanics and engineering
  195~(41) (2006) 5257--5296.

\bibitem{hu2017skew}
Q.~Hu, F.~Chouly, G.~Cheng, S.~P.~A. Bordas, Skew-symmetric nitsche's
  formulation in isogeometric analysis: Dirichlet and symmetry conditions,
  patch coupling and frictionless contact, arXiv preprint arXiv:1711.10253v2.

\bibitem{kiendl2010bending}
J.~Kiendl, Y.~Bazilevs, M.-C. Hsu, R.~W{\"u}chner, K.-U. Bletzinger, The
  bending strip method for isogeometric analysis of {K}irchhoff--{L}ove shell
  structures comprised of multiple patches, Computer Methods in Applied
  Mechanics and Engineering 199~(37) (2010) 2403--2416.

\bibitem{echter2010numerical}
R.~Echter, M.~Bischoff, Numerical efficiency, locking and unlocking of {NURBS}
  finite elements, Computer Methods in Applied Mechanics and Engineering
  199~(5) (2010) 374--382.

\bibitem{hughes1978reduced}
T.~J. Hughes, M.~Cohen, M.~Haroun, Reduced and selective integration techniques
  in the finite element analysis of plates, Nuclear Engineering and Design
  46~(1) (1978) 203--222.

\bibitem{adam2014improved}
C.~Adam, S.~Bouabdallah, M.~Zarroug, H.~Maitournam, Improved numerical
  integration for locking treatment in isogeometric structural elements, {P}art
  {I}: Beams, Computer Methods in Applied Mechanics and Engineering 279 (2014)
  1 -- 28.

\bibitem{adam2015improved}
C.~Adam, S.~Bouabdallah, M.~Zarroug, H.~Maitournam, Improved numerical
  integration for locking treatment in isogeometric structural elements. {P}art
  {II}: Plates and shells, Computer Methods in Applied Mechanics and
  Engineering 284 (2015) 106--137.

\bibitem{adam2015reduced}
C.~Adam, S.~Bouabdallah, M.~Zarroug, H.~Maitournam, {A Reduced Integration for
  Reissner-Mindlin Non-linear Shell Analysis Using T-Splines}, in: Isogeometric
  Analysis and Applications 2014, Springer, 2015, pp. 103--125.

\bibitem{Elguedj_2008_BoverbarFover_2732_2762}
T.~Elguedj, Y.~Bazilevs, V.~M. Calo, T.~J.~R. Hughes, B over-bar and {F}
  over-bar projection methods for nearly incompressible linear and non-linear
  elasticity and plasticity using higher-order nurbs elements, Computer Methods
  in Applied Mechanics and Engineering 197~(33-40) (2008) 2732--2762.

\bibitem{Bouclier_2012_Lockingfreeisogeometric_144_162}
R.~Bouclier, T.~Elguedj, A.~Combescure, Locking free isogeometric formulations
  of curved thick beams, Computer Methods in Applied Mechanics and Engineering
  245 (2012) 144--162.

\bibitem{bouclier2013development}
R.~Bouclier, T.~Elguedj, A.~Combescure, On the development of {NURBS}-based
  isogeometric solid shell elements: 2{D} problems and preliminary extension to
  3{D}, Computational Mechanics 52~(5) (2013) 1085--1112.

\bibitem{bouclier2013efficient}
R.~Bouclier, T.~Elguedj, A.~Combescure, Efficient isogeometric {NURBS}-based
  solid-shell elements: Mixed formulation and {B}-bar method, Computer Methods
  in Applied Mechanics and Engineering 267 (2013) 86--110.

\bibitem{bouclier2015isogeometric}
R.~Bouclier, T.~Elguedj, A.~Combescure, An isogeometric locking-free
  nurbs-based solid-shell element for geometrically nonlinear analysis,
  International Journal for Numerical Methods in Engineering 101~(10) (2015)
  774--808.

\bibitem{Echter_2013_hierarchicfamilyisogeometric_170_180}
R.~Echter, B.~Oesterle, M.~Bischoff, A hierarchic family of isogeometric shell
  finite elements, Computer Methods in Applied Mechanics and Engineering 254
  (2013) 170--180.

\bibitem{bletzinger2000unified}
K.-U. Bletzinger, M.~Bischoff, E.~Ramm, A unified approach for
  shear-locking-free triangular and rectangular shell finite elements,
  Computers \& Structures 75~(3) (2000) 321--334.

\bibitem{brezzi2010new}
F.~Brezzi, J.~Evans, T.~Hughes, L.~Marini, New quadrilateral plate elements
  based on {T}wist-{K}irchhoff theory, Comput Methods Appl Mech Eng
  (submitted).

\bibitem{brezzi2013virtual}
F.~Brezzi, L.~D. Marini, Virtual element methods for plate bending problems,
  Computer Methods in Applied Mechanics and Engineering 253 (2013) 455--462.

\bibitem{da2012avoiding}
L.~B. da~Veiga, C.~Lovadina, A.~Reali, Avoiding shear locking for the
  {T}imoshenko beam problem via isogeometric collocation methods, Computer
  Methods in Applied Mechanics and Engineering 241 (2012) 38--51.

\bibitem{auricchio2013locking}
F.~Auricchio, L.~B. da~Veiga, J.~Kiendl, C.~Lovadina, A.~Reali, Locking-free
  isogeometric collocation methods for spatial {T}imoshenko rods, Computer
  Methods in Applied Mechanics and Engineering 263 (2013) 113--126.

\bibitem{yin2014isogeometric}
S.~Yin, J.~S. Hale, T.~Yu, T.~Q. Bui, S.~P. Bordas, Isogeometric locking-free
  plate element: a simple first order shear deformation theory for functionally
  graded plates, Composite Structures 118 (2014) 121--138.

\bibitem{kiendl2015single}
J.~Kiendl, F.~Auricchio, T.~Hughes, A.~Reali, Single-variable formulations and
  isogeometric discretizations for shear deformable beams, Computer Methods in
  Applied Mechanics and Engineering 284 (2015) 988--1004.

\bibitem{beirao2015locking}
L.~Beir{\~a}o Da~Veiga, T.~Hughes, J.~Kiendl, C.~Lovadina, J.~Niiranen,
  A.~Reali, H.~Speleers, A locking-free model for {R}eissner--{M}indlin plates:
  {A}nalysis and isogeometric implementation via {NURBS} and triangular
  {NURPS}, Mathematical Models and Methods in Applied Sciences 25~(08) (2015)
  1519--1551.

\bibitem{hu2016order}
P.~Hu, Q.~Hu, Y.~Xia, Order reduction method for locking free isogeometric
  analysis of {T}imoshenko beams, Computer Methods in Applied Mechanics and
  Engineering 308 (2016) 1--22.

\bibitem{mitchell2011method}
T.~J. Mitchell, S.~Govindjee, R.~L. Taylor, A method for enforcement of
  {Dirichlet} boundary conditions in isogeometric analysis, in: Recent
  Developments and Innovative Applications in Computational Mechanics,
  Springer, 2011, pp. 283--293.

\bibitem{govindjee2012convergence}
S.~Govindjee, J.~Strain, T.~J. Mitchell, R.~L. Taylor, Convergence of an
  efficient local least-squares fitting method for bases with compact support,
  Computer Methods in Applied Mechanics and Engineering 213 (2012) 84--92.

\bibitem{nguyen2015isogeometric}
V.~P. Nguyen, C.~Anitescu, S.~P. Bordas, T.~Rabczuk, Isogeometric analysis: an
  overview and computer implementation aspects, Mathematics and Computers in
  Simulation 117 (2015) 89--116.

\bibitem{koschnick2005discrete}
F.~Koschnick, M.~Bischoff, N.~Camprub{\'\i}, K.-U. Bletzinger, The discrete
  strain gap method and membrane locking, Computer Methods in Applied Mechanics
  and Engineering 194~(21) (2005) 2444--2463.

\bibitem{simo1986variational}
J.~Simo, T.~Hughes, On the variational foundations of assumed strain methods,
  Journal of Applied Mechanics 53~(1) (1986) 51--54.

\bibitem{huang1994quasi}
B.-Z. Huang, V.~B. Shenoy, S.~Atluri, A quasi-conforming triangular laminated
  composite shell element based on a refined first-order theory, Computational
  mechanics 13~(4) (1994) 295--314.

\bibitem{Hughes1980}
T.~J.~R. Hughes, {Generalization of selective integration procedures to
  anisotropic and nonlinear media}, {Internat. J. Numer. Methods Engrg.} {15}
  ({1980}) {1413--1418}.

\bibitem{antolin2016isogeometric}
P.~Antolin, A.~Bressan, A.~Buffa, G.~Sangalli, An isogeometric method for
  linear nearly-incompressible elasticity with local stress projection,
  Computer Methods in Applied Mechanics and Engineering.

\bibitem{jlmmz2014}
B.~Juettler, U.~Langer, A.~Mantzaflaris, S.~Moore, W.~Zulehner, Geometry +
  simulation modules: Implementing isogeometric analysis, Proc. Appl. Math.
  Mech. 14~(1) (2014) 961--962.

\bibitem{macneal1985proposed}
R.~H. Macneal, R.~L. Harder, A proposed standard set of problems to test finite
  element accuracy, Finite elements in analysis and design 1~(1) (1985) 3--20.

\bibitem{bordas2010strain}
S.~P. Bordas, T.~Rabczuk, N.-X. Hung, V.~P. Nguyen, S.~Natarajan, T.~Bog, N.~V.
  Hiep, et~al., Strain smoothing in {FEM} and {XFEM}, Computers \& structures
  88~(23) (2010) 1419--1443.

\bibitem{chen2012extended}
L.~Chen, T.~Rabczuk, S.~P.~A. Bordas, G.~Liu, K.~Zeng, P.~Kerfriden, Extended
  finite element method with edge-based strain smoothing ({ES}m-{XFEM}) for
  linear elastic crack growth, Computer Methods in Applied Mechanics and
  Engineering 209 (2012) 250--265.

\bibitem{surendran2017linear}
M.~Surendran, S.~Natarajan, S.~Bordas, G.~Palani, Linear smoothed extended
  finite element method, arXiv preprint arXiv:1701.03997.

\end{thebibliography}

\end{document}